\documentclass[10pt]{article}
\usepackage{iclr2027_conference}
\iclrfinalcopy
\usepackage{fontspec}
\usepackage{amsmath,amssymb,xcolor,xspace,enumitem,graphicx,booktabs,pifont}

\newcommand{\ours}{SpheraAudio\xspace}

\newsavebox{\reviewtablebox}
\makeatletter
\newenvironment{reviewtable}{\par\smallskip\noindent\begin{minipage}{\linewidth}\def\@captype{table}\setlength{\abovecaptionskip}{0pt}\centering}{\end{minipage}\par\smallskip}
\makeatother

\title{Any Channel Count, One Token Stream: Incorporating Spatial Audio Understanding and Reasoning into Large Audio-Language Models}
\author{Yuhuan You$^{1}$, Wenxiang Guo$^{2}$, Zhifang Guo$^{1}$,\\
Yuxuan Wang$^{1}$, Jin Xu$^{1}$\\
$^{1}$Alibaba Group \qquad $^{2}$Zhejiang University}
\usepackage{tcolorbox}
\tcbuselibrary{breakable}
\definecolor{SpheraInk}{HTML}{243744}
\definecolor{SpheraBorder}{HTML}{CBD5DC}
\definecolor{SpheraWash}{HTML}{F5F8FA}
\definecolor{SpheraAccent}{HTML}{236F7A}
\definecolor{SpheraMuted}{HTML}{56626D}
\newtcolorbox{spheracase}[2][]{%
  colback=white,colframe=SpheraBorder,colbacktitle=SpheraWash,
  coltitle=SpheraInk,fonttitle=\small\bfseries,fontupper=\small,
  title={#2},boxrule=.45pt,arc=3pt,outer arc=3pt,boxsep=0pt,
  left=7pt,right=7pt,top=6pt,bottom=6pt,toptitle=4pt,bottomtitle=4pt,
  before skip=7pt,after skip=7pt,before upper={\raggedright\setlength{\parindent}{0pt}\setlength{\parskip}{2pt}},
  #1}
\newcommand{\spherafield}[2]{\noindent{\bfseries\color{SpheraInk}#1}\enspace#2\par}
\newcommand{\spheradivider}{\par\vspace{3pt}{\color{SpheraBorder}\hrule height .35pt}\vspace{5pt}}

\newcommand{\spheraillustratedquery}[2]{%
  \noindent\begin{minipage}[c]{.235\linewidth}
    \centering\includegraphics[width=\linewidth]{#1}
  \end{minipage}\hfill
  \begin{minipage}[c]{.735\linewidth}\raggedright#2\end{minipage}\par}

\newcommand{\spherarow}[2]{%
  \par\smallskip\begingroup\hangindent=.105\linewidth\hangafter=1
  \noindent\makebox[.105\linewidth][l]{{\bfseries\color{SpheraInk}#1}}#2\par\endgroup}

\begin{document}
\raggedbottom
\maketitle
\fancyhead{}
\renewcommand{\headrulewidth}{0pt}
\begin{abstract}
Audio carries phase relations that encode where sources are, yet existing large audio-language models (LALM) reduce it to single-channel content for stability of general audio understanding, leaving essential spatial information for answering questions about direction, distance, and source relations unused. We present SpheraAudio, which accepts recordings with any channel count --- mono, stereo, First-Order Ambisonics (FOA), or higher --- and maps them into a single token stream. A spatial encoder turns complex multichannel spectra into spatial tokens that are interleaved with audio tokens at corresponding temporal positions, while we propose Spatial Multimodal RoPE (SMRoPE), which makes each token's modality identity and temporal slot visible to rotary attention without altering the language model's original RoPE. Training runs in three stages --- spatial-encoder pretraining, projector alignment, and joint supervised-and-reinforcement learning --- progressing from geometric supervision to language-based spatial decisions, with answers trained as an explicit spatial chain-of-thought: step-by-step geometric reasoning over the heard scene. Considering that current benchmarks tie their questions to a fixed channel-count format and score answers without spatial reasoning evidence, we introduce SpheraQA, which spans mono, stereo, and FOA recordings and organises questions hierarchically from single-attribute perception through cross-source relations to multi-step reasoning, scored programmatically against simulated scene geometry and real-recording annotations with step-by-step solution references. Across SpheraQA and public spatial-audio benchmarks, SpheraAudio sets a new performance standard for spatial audio understanding. On spatial tasks, it supports high-precision localisation, tracking, and reasoning over static and moving sound sources; beyond them, the model preserves strong general-audio capabilities. We will release the code, the SpheraQA benchmark, and its evaluation metrics to support reproducibility.

\end{abstract}
\section{Introduction}
\label{sec:introduction}

Binaural hearing helps humans locate sound sources and follow their motion, providing spatial information that complements vision. Bringing this spatial awareness to large audio-language models (LALMs) can support perception, navigation, and interaction in embodied agents operating in dynamic environments.

Traditional sound source localization and sound event localization and detection (SELD) methods estimate source directions, positions, and activity~\citep{seldnet,pseldnets}. Their outputs primarily consist of task-specific estimates, such as coordinates and event labels, leaving open-ended scene interpretation and spatial reasoning to downstream systems. Recent LALMs support language-based audio understanding, but many general-purpose models retain mono-audio processing to maintain stable general-audio performance and limit potential interference from redundant or task-irrelevant information in multichannel inputs, leaving inter-channel spatial cues unused~\citep{qwen25omni}.

Recent spatial extensions address this gap but differ in their input requirements and feature integration. BAT is designed for two-channel binaural audio~\citep{bat}. Spatial-Omni-7B is a recent spatial-audio LLM that takes an initial step toward the LLM paradigm for spatial-audio training and evaluation, but does not systematically examine its effects on single-channel general-audio capabilities~\citep{spatialomni}. PhaseCoder introduces a spatial encoder for microphone-array recordings with varying channel counts, yet requires microphone coordinates; when integrated with a LALM, it prepends spatial tokens to the audio token sequence, leaving temporal correspondence to be learned implicitly during fine-tuning~\citep{phasecoder}. Furthermore, spatial adaptation risks degrading general-audio performance, and findings from low-rank adaptation (LoRA) or other lightweight adapters cannot be directly extrapolated to full-parameter updates. We therefore investigate explicit temporal pairing alongside full-parameter spatial adaptation, evaluating spatial understanding together with general-audio performance.

We present SpheraAudio, a LALM framework for understanding and reasoning about spatial sound scenes. SpheraAudio comprises an audio encoder, Audio Transformer (AuT), and a spatial encoder, Spatial~AuT. AuT extracts audio-content features, while Spatial~AuT extracts spatial features from complex multichannel spectra. Their respective projectors map these features into audio tokens and spatial tokens. We interleave time-aligned audio tokens and spatial tokens into a shared sequence for the language model, using Spatial Multimodal Rotary Position Embedding (SMRoPE) to encode their temporal correspondence and modality identities within rotary attention. Through appropriate input mappings, the framework supports mono, stereo, first-order Ambisonics (FOA), and other compatible multichannel recordings, including higher-order Ambisonics (HOA) and calibrated microphone-array inputs.

Training begins with geometric pretraining of Spatial~AuT, followed by projector alignment within joint supervised fine-tuning (SFT). This stage emphasizes explicit, fine-grained spatial audio chain-of-thought (CoT) supervision: the targets specify source identities, time intervals, spatial attributes, and geometric operations, teaching the model to associate audible sources with spatial observations and derive answers step by step. Finally, outcome-based reinforcement learning (RL) refines spatial decisions using three types of final-answer rewards: numerical proximity, correctness of judgments or choices, and hitting-list matching.

Existing spatial-audio benchmarks differ in recording formats, task coverage, and scoring criteria, complicating a unified assessment of geometric precision and reasoning across sources and time. We therefore introduce SpheraQA, a benchmark containing $20{,}492$ questions organized into three levels: L1 perception, L2 relations, and L3 reasoning. Its tasks progress from identifying individual attributes, such as direction and distance, to describing trajectories and combining observations across sources and time. Numerical, categorical, and structured answers are scored against simulated scene geometry and real-recording annotations using explicit task-specific criteria. Step-by-step solution references additionally support inspection of the geometric reasoning underlying an answer. This design connects language-based evaluation with concrete spatial measurements and scene relationships.

At the four-billion-parameter (4B) scale, SpheraAudio achieves state-of-the-art performance on the public SO-Bench and TWNM benchmarks, outperforming Spatial-Omni-7B, and obtains the strongest overall results on SpheraQA. Its gains in precise localization, distance-change reasoning, and source-identity binding connect spatial perception to geometric reasoning. Dedicated stereo experiments demonstrate two-channel laterality and motion understanding, while controlled sound-field transformations show that spatial predictions track changes in directional input. A general-audio study further shows improvements over mono-only adaptation, supporting spatial learning alongside retained general-audio capabilities.

\label{end:introduction}

\section{Related Work}

\label{sec:previous_work}

\paragraph{Sound source localization.} Localization estimates directions from inter-channel timing and phase; SELD pairs positions with activity and event labels~\citep{knapp1976gcc,seldnet,pseldnets}. Geometry-conditioned methods generalize across microphone layouts~\citep{grinstein2023neuralsrp,baek2025gidoaenet}. Spatial language queries additionally require binding sound identity to geometry across observations.

\paragraph{Spatial audio-language representations.} ELSA learns contrastive audio--text embeddings that combine sound semantics with spatial attributes for retrieval and localization~\citep{elsa2024}. Generative models impose more specific input assumptions: BAT and OWL specialize in binaural recordings~\citep{bat,owl}, while SPUR and Spatial-Omni use FOA~\citep{spur,spatialomni}. Spatial-Omni-7B advances spatial-audio training and evaluation toward the LLM paradigm, yet its effects on single-channel general-audio capabilities remain systematically unexplored. PhaseCoder conditions on microphone coordinates to accommodate varying array layouts~\citep{phasecoder}.

Channel compatibility also depends on device acoustics. PhaseCoder assumes free-floating omnidirectional microphones, whereas deployed devices introduce housing and baffle scattering, microphone directivity, and mounting effects. Identical microphone coordinates can thus yield distinct direction- and frequency-dependent responses. Calibrated array-to-Ambisonics encoding incorporates measured device responses into a shared FOA basis~\citep{gayer2024ambisonics,zotter2019ambisonics}. SpheraAudio adopts this interface for compatible arrays, alongside stereo mid--side mapping and first-order extraction from HOA.

Content must also be associated with spatial evidence. Sci-Phi and PhaseCoder separate spatial and audio token blocks~\citep{jiang2025sciphi,phasecoder}; ST-AudioLM pairs a clip-level semantic summary with trajectory-aware tokens~\citep{stAudioLM}. SpheraAudio instead establishes temporal pairs and encodes their correspondence with SMRoPE.

\paragraph{Spatial reasoning and model adaptation.} LTU and Audio Flamingo 2 use instruction tuning and reasoning curricula~\citep{ltu,audioflamingo2}; SARI and R1-AQA explore outcome rewards~\citep{sari,r1aqa}. OWL supervises source-relation explanations, and TWNM employs Soft Adaptive Policy Optimization (SAPO)~\citep{owl,twnm}. Spatial adaptation commonly uses LoRA or lightweight adapters~\citep{lora,owl,spur,spatialomni,phasecoder}, leaving full-parameter effects less explored. We study full-parameter adaptation with fine-grained spatial CoT and Group Relative Policy Optimization (GRPO)~\citep{grpo}, evaluating both spatial gains and general-audio performance.

\paragraph{Spatial audio benchmarks.} SpatialSoundQA and BiDepth test discretized binaural attributes and relations~\citep{bat,owl}; TWNM tests scene-level multiple-choice reasoning~\citep{twnm}, and SO-Bench covers localization, motion, relations, and semantics~\citep{spatialomni}. SpheraQA unifies perception, relations, and reasoning, requiring numerical coordinates, time-indexed trajectories, and geometric decisions to satisfy all queried components and timestamps.

\label{end:previous_work}

\section{Architecture}

\label{sec:architecture}

\begin{figure}[!ht]
\centering
\includegraphics[width=\textwidth]{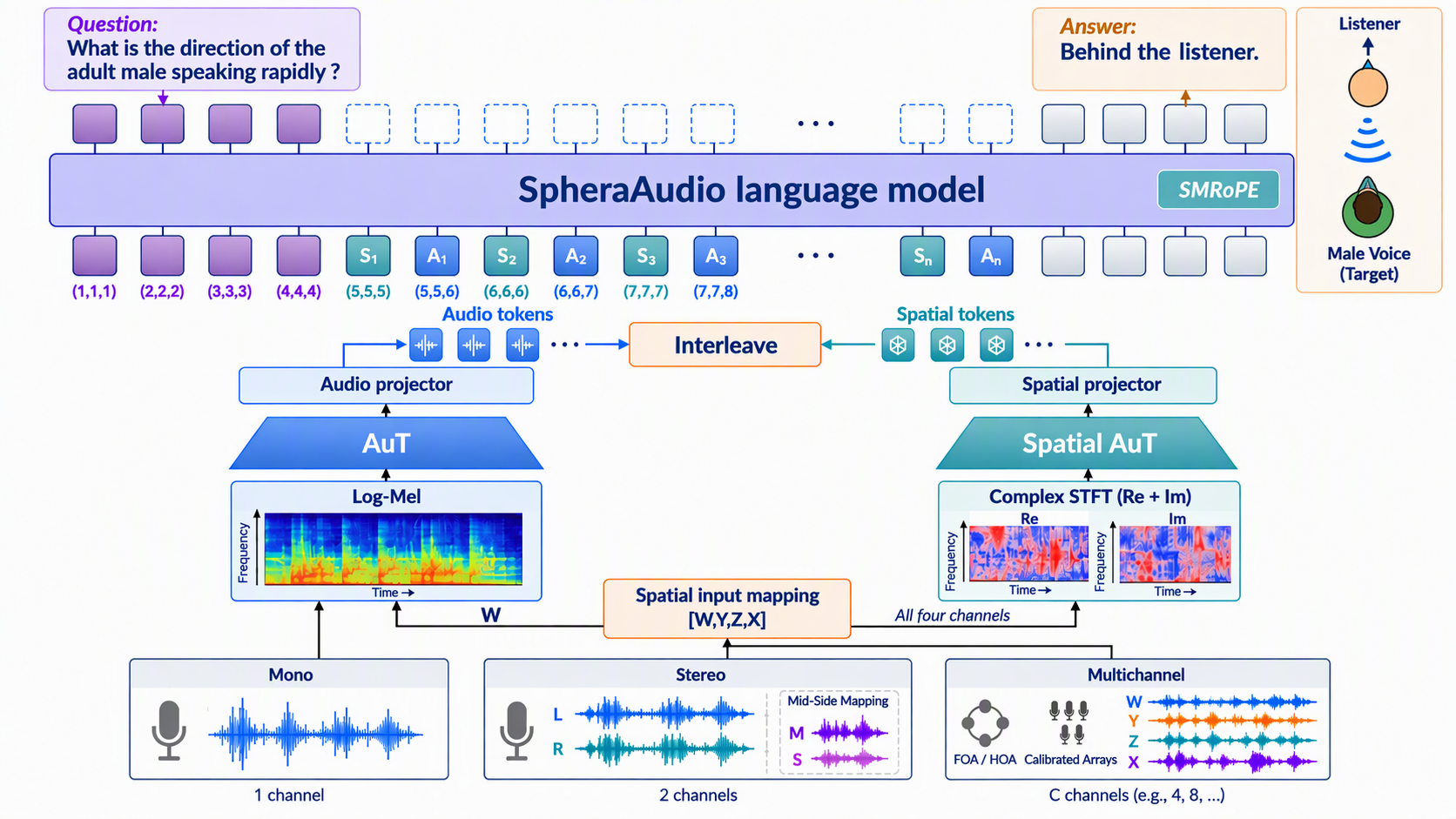}
\caption{\textbf{\ours overview.} The audio encoder AuT and spatial encoder Spatial AuT extract complementary features from the same recording. Separate projectors produce audio tokens and spatial tokens, which are paired in time and interleaved. SMRoPE encodes their temporal correspondence and modality identities for the language model.}
\label{fig:system}
\end{figure}

\ours combines the audio encoder AuT, spatial encoder Spatial~AuT, separate projectors, and a large language model (LLM). The input mappings use shared encoders across recording formats. Projected audio and spatial tokens are paired in time and interleaved, with SMRoPE encoding time and modality (Figure~\ref{fig:system}).

\subsection{Audio and Spatial Encoders}

\label{sec:spatialenc}

AuT extracts content features from a single-channel log-mel spectrogram. Spatial~AuT extracts spatial features from the real and imaginary components of multichannel short-time Fourier transform (STFT) coefficients. Appendix~\ref{app:spatialenc_layers} details the convolutional and temporal-attention architecture.

Let $\mathbf h^a_t$ and $\mathbf h^s_t$ denote the audio and spatial encoder features at temporal position $t$. The audio projector $P_a$ and spatial projector $P_s$ map these features into the language embedding space:
\begin{equation}
A_t=P_a(\mathbf h^a_t),\qquad S_t=P_s(\mathbf h^s_t).
\label{eq:token_definition}
\end{equation}
The projected embeddings $A_t$ and $S_t$ denote the \emph{audio token} and \emph{spatial token}, respectively, each matching the language embedding dimension of $2560$.

\subsection{Explicit Temporal Pairing across Modalities}

\label{sec:integration}

The two encoders share the input frame grid, chunk boundaries, and temporal downsampling schedule, yielding temporally aligned features and projected tokens (Appendix~\ref{app:input_features}).

We place each spatial token immediately before its corresponding audio token. For $K$ paired output positions, the audio segment of the language-model input is
\begin{equation}
\mathcal{X}
=
[\langle\mathrm{audio}\rangle,
S_1,A_1,\ldots,S_K,A_K,
\langle/\mathrm{audio}\rangle].
\label{eq:paired_sequence}
\end{equation}
Under the causal attention mask, this ordering allows $A_t$ to attend to its paired $S_t$, while the reverse same-step interaction remains masked. The two token positions retain separate key and value representations, allowing the language model to attend to audio content and spatial information individually.

SMRoPE adapts rotary position embeddings (RoPE) via coordinates inspired by multimodal RoPE~\citep{qwen25omni}. Like neighboring visual patches in one frame, paired spatial and audio tokens share temporal and row coordinates but differ in the third coordinate, retaining the backbone's rotary operation.

With positional offset $b$ before the first paired time step, SMRoPE assigns
\begin{equation}
\mathbf p(S_t)=(b+t,b+t,b+t),
\qquad
\mathbf p(A_t)=(b+t,b+t,b+t+1).
\label{eq:smrope}
\end{equation}
The shared baseline advances with acoustic time, while the third coordinate distinguishes modalities. Within a modality, relative positions follow temporal displacement; between modalities, the third coordinate adds a signed phase offset. Appendix~\ref{app:smrope_derivation} derives the key-minus-query coordinates and attention kernel, including frequency allocation and boundary-token positions.

\subsection{Shared Representations across Channel Counts}
\label{sec:foa_interface}

Input mappings maintain a common four-component spatial input. FOA contains the omnidirectional $W$ and directional components in Ambisonic Channel Number (ACN) order $[W,Y,Z,X]$, using three-dimensional Schmidt semi-normalization (SN3D)~\citep{zotter2019ambisonics}. Native FOA enters directly; HOA supplies its zeroth- and first-order components (Appendix~\ref{app:input_features}).

Microphone directivity, mounting, and scattering alter device transfer functions beyond what array coordinates capture. Calibrated array-to-Ambisonics encoding incorporates these measured responses into the shared spatial basis~\citep{phasecoder,mccormack2022parametric,gayer2024ambisonics}:
\begin{equation}
\widehat{\mathbf a}_1(f,t)=\mathbf E_C(f)\mathbf x_C(f,t),\qquad \mathbf E_C(f)\in\mathbb C^{4\times C}.
\end{equation}
Here $\mathbf x_C(f,t)$ is the STFT vector of $C$ microphone signals at frequency $f$ and time $t$. Calibration determines $\mathbf E_C(f)$ from measured directional responses, allowing one spatial representation to serve compatible capture devices.

Stereo uses the invertible mid--side mapping
\begin{equation}
W_m=\frac{L+R}{2},\quad Y_m=\frac{L-R}{2},\quad \widetilde{\mathbf a}=[W_m,Y_m,0,0]^{\mathsf T}.
\label{eq:stereo_lift}
\end{equation}
The inverse $L=W_m+Y_m$, $R=W_m-Y_m$ preserves both waveforms. Spatial AuT processes the complex spectra of $\widetilde{\mathbf a}$, while AuT receives $W_m$. Mono retains the base model's native AuT-only sequence and positions.\label{end:architecture}

\section{Training}
\label{sec:training}

\begin{figure}[!ht]
\centering
\includegraphics[width=0.96\textwidth]{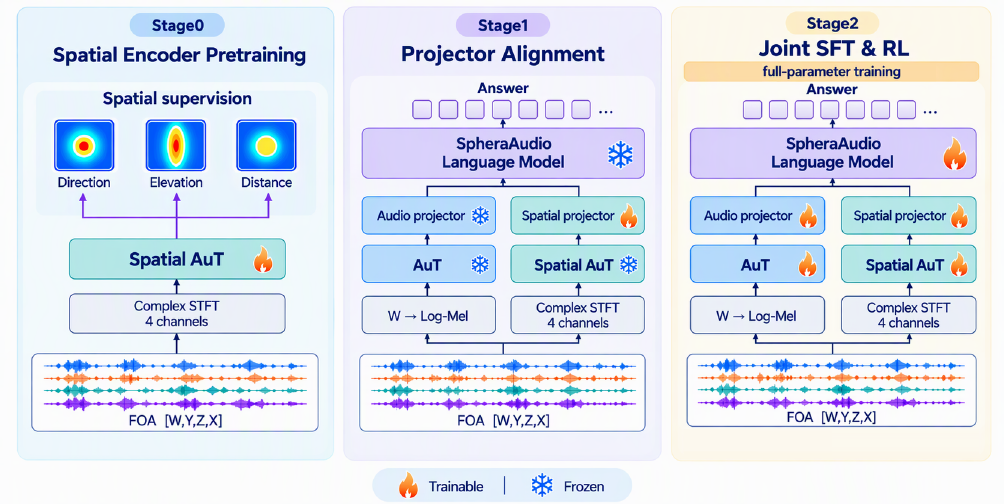}
\caption{\textbf{Schematic training framework.} Spatial-encoder pretraining learns temporally resolved geometry. Projector alignment connects spatial features to the pretrained language representation. Joint SFT \& RL combines spatial and acoustic evidence through supervised answer learning and group-relative optimization with verifiable outcome rewards.}
\label{fig:training_stages}
\end{figure}

Training combines Spatial~AuT pretraining and projector alignment with joint SFT and outcome-based RL to learn audio-conditioned spatial decisions (Figure~\ref{fig:training_stages}).

\subsection{Spatial-Encoder Pretraining}
\label{sec:stage1}

We pretrain Spatial~AuT from scratch on simulated FOA with temporal source annotations and room properties. Auxiliary heads predict azimuth, elevation, distance, activity, room floor area, and wall absorption. Regression uses mean squared error at active positions, with sine--cosine azimuth targets; activity uses binary cross-entropy over sound and silence. Appendix~\ref{app:spatial_pretraining} specifies the masked losses and task balancing.

\subsection{Joint SFT}
\label{sec:stage2}
\label{sec:projector_alignment}

Projector alignment first maps spatial features into the language embedding space with other modules frozen. Joint SFT subsequently mixes general-audio samples with spatial CoT supervision, optimizing both encoders, both projectors, and the language backbone via an autoregressive loss over reasoning traces and final answers.

Spatial CoT targets are constructed from each training recording's source annotations. They identify the queried source and time, state spatial observations, apply geometric operations, and derive the answer. Counting uses source activity and attributes; navigation derives a turn from direction; relations compare sources in a shared coordinate frame. The model receives audio and a question, with annotations supplying reasoning and answer supervision (Appendices~\ref{app:training_details} and~\ref{app:question_generation}).

\subsection{Reinforcement Learning}
\label{sec:stage3}

Outcome rewards follow three answer formats. Numerical answers receive tolerance-based partial credit from their reference error: circular error for azimuth and listener rotation, and absolute error for elevation and distance. Categorical and multiple-choice answers receive correctness rewards. Descriptions receive the weighted fraction of matched reference-checklist entries. Appendix~\ref{app:training_details} specifies the scoring functions.

A single RL stage applies GRPO~\citep{grpo}, converting final-answer rewards across sampled responses into group-normalized advantages. RL updates both encoders, both projectors, and the language backbone; the final SFT policy initializes training and supplies the fixed Kullback--Leibler (KL) reference. SFT supervises spatial observations and operations, while RL rewards final answers.

\section{Evaluation}
\label{sec:evaluation}

We evaluate spatial reasoning, general-audio retention, stereo compatibility, and the contributions of token integration and outcome rewards. The language backbone starts from the Qwen3-Omni 4B pretrained LLM~\citep{qwen3omni}, and Spatial AuT from geometric pretraining. Appendix~\ref{app:implementation} details architecture and optimization.

\subsection{Spatial Understanding and Reasoning}
\label{sec:eval_spatial}
\label{sec:spatialqa}
\label{sec:eval_llm}
\label{sec:eval_ood}

\begin{figure}[!ht]
\centering
\includegraphics[width=0.96\textwidth]{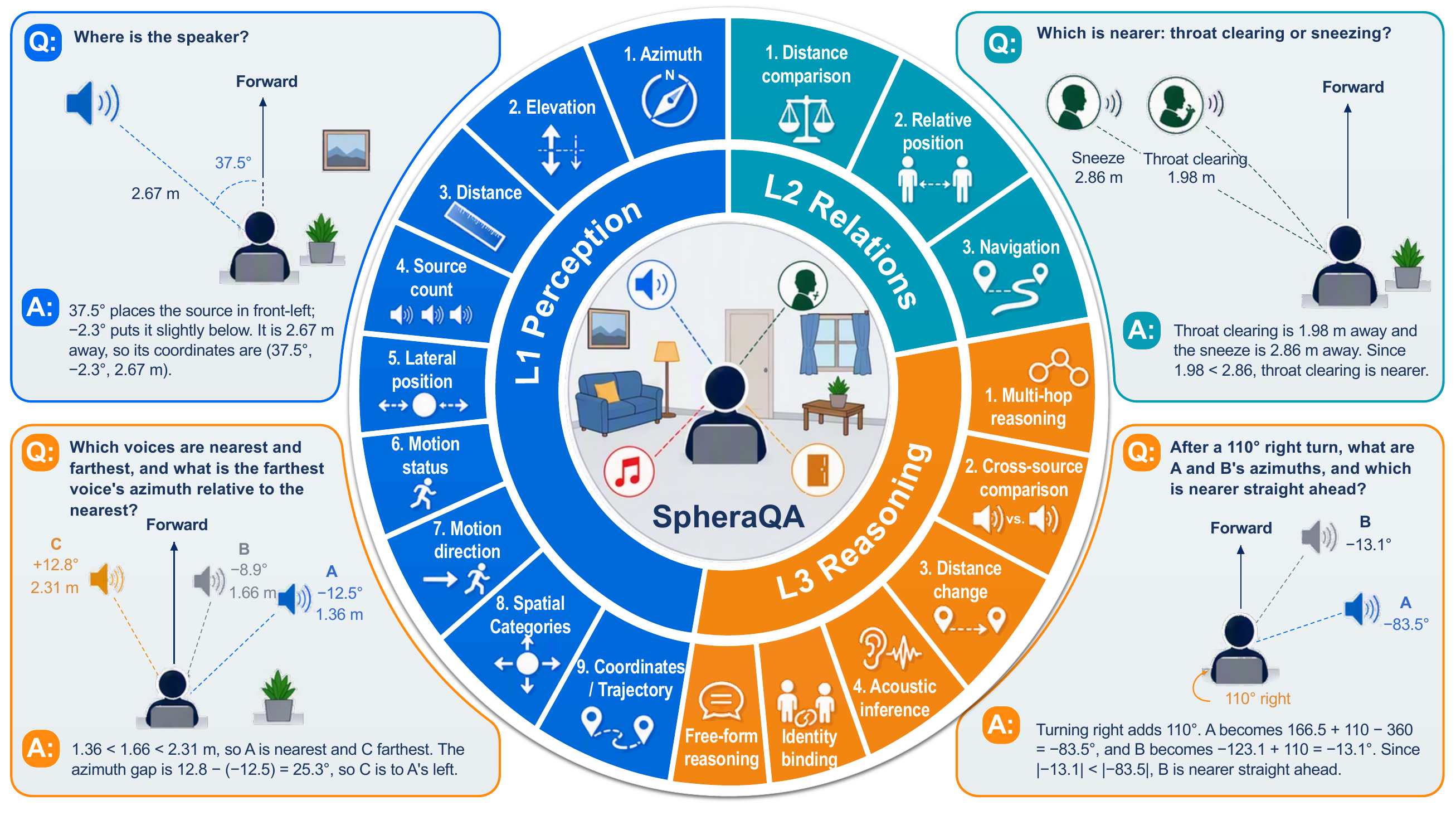}
\caption{\textbf{SpheraQA.} The task taxonomy spans single-attribute perception, cross-source relations, and spatial reasoning. Examples connect source observations to comparisons and geometric operations.}
\label{fig:benchmark_constellation}
\end{figure}

\textbf{SpheraQA: from attributes to spatial decisions.} SpheraQA contains $20{,}492$ questions spanning L1 perception, L2 relations, and L3 reasoning (Figure~\ref{fig:benchmark_constellation}). Tasks progress from source attributes to relations and geometric decisions across sources and time. Table~\ref{tab:spatial_main} reports task-level and overall whole-answer accuracy.

Numerical answers use geometric tolerances; coordinates and trajectories must satisfy every requested component and timestamp. GPT-4o extracts values and source references from free-form responses for programmatic scoring (Appendix~\ref{app:spatialqa}).

\textbf{Public spatial benchmarks.} SO-Bench~\citep{spatialomni} measures localization, motion, relations, and semantics via mean per-question scores. TWNM~\citep{twnm} tests perception, relations, and reasoning through multiple-choice questions, scored by option likelihood for SpheraAudio and Spatial-Omni-7B and generated letters for Gemini (Appendix~\ref{app:benchmarks}).

SpheraAudio and Spatial-Omni-7B~\citep{spatialomni} receive the same multichannel recordings and question prompts. Gemini 3.1 Pro receives only the omnidirectional $W$ channel, testing how far single-channel audio understanding supports these spatial questions (Table~\ref{tab:spatial_main}).

\begin{reviewtable}
\centering
\small
\caption{Spatial understanding (\%). SpheraQA reports whole-answer accuracy, with Spatial Reasoning pooling six reasoning question types. SO-Bench reports the mean per-question score. Gemini receives only $W$. $N$ counts questions and $\Delta$ is SpheraAudio minus Spatial-Omni-7B in percentage points before rounding. Best scores are in bold.}
\label{tab:spatial_main}
\label{tab:spatialqa_detail}
\setlength{\tabcolsep}{2pt}
\renewcommand{\arraystretch}{1.0}
\begin{tabular*}{\linewidth}{@{\extracolsep{\fill}}lrrrrr@{}}
\toprule
Task / benchmark & $N$ & \shortstack{Spatial-\\Omni-7B} & \shortstack{Gemini 3.1\\Pro} & SpheraAudio & $\Delta$ (pp) \\
\midrule
\multicolumn{6}{@{}l}{\textbf{(a) SpheraQA by question type}} \\
Azimuth & 1,563 & 22.39 & 9.72 & \textbf{55.60} & +33.21 \\
Elevation & 1,556 & 35.86 & 32.46 & \textbf{69.34} & +33.48 \\
Distance & 1,564 & 49.55 & 55.75 & \textbf{74.81} & +25.26 \\
Source count & 1,107 & 46.52 & 54.29 & \textbf{60.70} & +14.18 \\
Lateral position & 1,315 & 29.66 & \textbf{44.64} & 41.52 & +11.86 \\
Motion status & 1,604 & 63.34 & \textbf{73.94} & 73.38 & +10.04 \\
Motion direction & 1,314 & 45.36 & \textbf{49.39} & 34.93 & -10.43 \\
Spatial categories & 5,852 & 59.11 & 56.49 & \textbf{74.57} & +15.46 \\
Coordinates & 1,451 & 2.07 & 1.31 & \textbf{35.91} & +33.84 \\
Trajectory & 897 & 0.45 & 0.45 & \textbf{36.12} & +35.67 \\
Distance comparison & 334 & 32.93 & \textbf{37.43} & 32.04 & -0.90 \\
Relative position & 229 & 31.88 & \textbf{41.05} & 32.75 & +0.87 \\
Navigation & 1,586 & 40.04 & 39.28 & \textbf{58.83} & +18.79 \\
Spatial Reasoning & 120 & 19.17 & 27.50 & \textbf{30.00} & +10.83 \\
\midrule
\multicolumn{6}{@{}l}{\textbf{(b) Benchmark summaries}} \\
SpheraQA & 20,492 & 41.65 & 42.73 & \textbf{60.18} & +18.53 \\
SO-Bench & 7,877 & 64.61 & 32.32 & \textbf{66.94} & +2.32 \\
TWNM & 1,000 & 46.20 & 56.60 & \textbf{71.70} & +25.50 \\
\bottomrule
\end{tabular*}
\end{reviewtable}

SpheraAudio leads on all three spatial benchmarks and achieves $60.18\%$ overall accuracy on SpheraQA, exceeding Spatial-Omni-7B by $18.53$ percentage points (Table~\ref{tab:spatial_main}). Its largest gains arise in direction, coordinates, and trajectories. Spatial AuT pretraining targets these geometric quantities, while spatial CoT connects them to queried sources. The gain on Spatial Reasoning extends this connection from identifying spatial attributes to combining source identity and geometry across observations.

Gemini 3.1 Pro performs well on coarse lateral and motion judgments but trails SpheraAudio in precise azimuths and complete coordinates. Single-channel audio retains content and some distance or motion cues while omitting inter-channel directional evidence. These results support an explicit spatial pathway for fine-grained localization alongside audio-content understanding.

Appendix~\ref{app:astra_tools} also reports a $100$-question diagnostic with a waveform-analysis agent.

\subsection{General Audio Understanding}
\label{sec:eval_general}

A separate study examines general-audio retention following full-parameter adaptation by comparing Base, Adapted (Mono Data), and Adapted (Mono + Spatial Data). Base utilizes the Qwen3-Omni 4B pretrained LLM with its original audio pathway. The mixed-data configuration combines mono and spatial examples and incorporates the spatial pathway.

Transcription reports character error rate (CER) for Mandarin and word error rate (WER) for English~\citep{fleurs,commonvoice,wenetspeech}; emotion and vocal-sound classification report accuracy~\citep{iemocap,meld,vocalsound}. Table~\ref{tab:general_main} lists datasets; Appendix~\ref{app:general_protocol} details prompts and parsing.

\begin{reviewtable}
\centering
\small
\caption{General-audio recognition for Base, Adapted (Mono Data), and Adapted (Mono + Spatial Data). Arrows indicate the direction of improvement; the best result per subset is bold.}
\label{tab:general_main}
\sbox{\reviewtablebox}{%
\begin{tabular}{@{}llrrr@{}}
\toprule
Test subset & Metric & Base & \shortstack{Adapted\\{\fontsize{8}{9.5}\selectfont (Mono Data)}} & \shortstack{Adapted\\{\fontsize{8}{9.5}\selectfont (Mono + Spatial Data)}} \\
\midrule
FLEURS (cmn) & CER (\%) $\downarrow$ & 11.14 & 6.10 & \textbf{2.97} \\
FLEURS (en) & WER (\%) $\downarrow$ & 4.51 & 6.62 & \textbf{4.38} \\
Common Voice 15 (zh) & CER (\%) $\downarrow$ & \textbf{7.69} & 20.33 & 12.71 \\
Common Voice 15 (en) & WER (\%) $\downarrow$ & 11.22 & 16.58 & \textbf{10.76} \\
WenetSpeech (meeting) & CER (\%) $\downarrow$ & 15.35 & 15.31 & \textbf{8.06} \\
WenetSpeech (net) & CER (\%) $\downarrow$ & 26.92 & 12.65 & \textbf{7.48} \\
IEMOCAP (session 5) & Acc. (\%) $\uparrow$ & 4.54 & 59.61 & \textbf{69.59} \\
MELD (test) & Acc. (\%) $\uparrow$ & 0.77 & 45.59 & \textbf{52.72} \\
VocalSound (test) & Acc. (\%) $\uparrow$ & 51.66 & 85.80 & \textbf{87.89} \\
\bottomrule
\end{tabular}%
}
\ifdim\wd\reviewtablebox>\linewidth
\resizebox{\linewidth}{!}{\usebox{\reviewtablebox}}
\else\usebox{\reviewtablebox}\fi
\end{reviewtable}

Adapted (Mono + Spatial Data) outperforms Adapted (Mono Data) on all nine subsets and Base on eight. General-audio examples maintain supervision for sound content, while mono inputs retain the native AuT-only sequence. SMRoPE preserves within-modality temporal displacements when spatial tokens are introduced. These training and input choices support spatial learning alongside retained or improved general-audio performance.

\subsection{Stereo Understanding}
\label{sec:eval_stereo}

To test two-channel compatibility, Table~\ref{tab:stereo_main} compares SpheraAudio after dedicated stereo SFT with BAT~\citep{bat} and Gemini on $416$ questions about laterality and motion. SpheraAudio uses the invertible mid--side mapping; Gemini receives only $(L+R)/2$. Appendix~\ref{app:native_stereo_results} separately reports transfer of the spatial SFT and RL policies.

\begin{reviewtable}
\centering
\small
\caption{Stereo question-answering accuracy (\%) across $416$ questions. SpheraAudio uses stereo SFT; Gemini 3.1 Pro receives only the mono mid channel $(L+R)/2$ through its native audio interface.}
\label{tab:stereo_main}
\sbox{\reviewtablebox}{%
\begin{tabular}{@{}lrrr@{}}
\toprule
Method & \shortstack{Left--right Acc. (\%)\\$N=71$} & \shortstack{Lateral position Acc. (\%)\\$N=73$} & \shortstack{Motion Acc. (\%)\\$N=272$} \\
\midrule
BAT~\citep{bat} & 70.4 & 68.5 & 55.9 \\
Gemini 3.1 Pro & 46.5 & 19.2 & 50.4 \\
SpheraAudio (stereo SFT) & \textbf{95.8} & \textbf{93.2} & \textbf{85.3} \\
\bottomrule
\end{tabular}%
}
\ifdim\wd\reviewtablebox>\linewidth
\resizebox{\linewidth}{!}{\usebox{\reviewtablebox}}
\else\usebox{\reviewtablebox}\fi
\end{reviewtable}

Stereo-adapted SpheraAudio achieves the highest performance across all three tasks. The invertible mapping preserves both recorded waveforms, enabling Spatial AuT to exploit inter-channel differences through the shared encoder and token interfaces.

\subsection{Ablations}
\label{sec:eval_ablation}

We evaluate token fusion, cross-stream temporal correspondence, and the coordinate grid used to retrieve temporal evidence.

\textbf{Separate token positions.} Element-wise addition degrades spatial understanding and general-audio transcription relative to interleaving (Table~\ref{tab:design_ablation}a). Interleaving preserves distinct projected audio and spatial representations for attention to query, improving angular and distance estimates, AISHELL transcription, and laterality and motion-state judgments in single-source scenes, with little change in trajectory accuracy.

\textbf{Temporal correspondence.} Named-source localization requires linking acoustic content to its spatial evidence. Shifting only the spatial token vectors reduces accuracy from 75.00\% to 25.00\%, whereas shifting both streams jointly maintains 81.25\% accuracy (Table~\ref{tab:design_ablation}b). A joint shift preserves audio--spatial pair associations, whereas a spatial-only shift misaligns acoustic content with geometry from a different time step. This contrast supports the explicit correspondence established in Section~\ref{sec:integration}.

\begin{reviewtable}
\centering
\small
\caption{Component ablations. (a) Token fusion in single-source scenes: spatial mean absolute error (MAE), AISHELL CER, and task accuracy. (b) Named-source direction under token-stream shifts. (c) Timed and untimed queries under SMRoPE and sequential RoPE. Panels (b,c) evaluate a fixed SFT checkpoint on identical confirmation recordings, with $N$ questions per task (Appendix~\ref{app:temporal_ablations}).}
\label{tab:design_ablation}
\setlength{\tabcolsep}{1.5pt}
\renewcommand{\arraystretch}{1.08}
\begin{minipage}[t]{1.0\linewidth}
\centering
\setlength{\tabcolsep}{2pt}
\textbf{(a) Token fusion}\par\vspace{3pt}
\begin{tabular*}{\linewidth}{@{\extracolsep{\fill}}lrrrrrrr@{}}
\toprule
 & \multicolumn{3}{c}{Spatial MAE$\downarrow$} & \multicolumn{1}{c}{Speech} & \multicolumn{3}{c}{Spatial Acc. (\%)$\uparrow$} \\
\cmidrule(lr){2-4} \cmidrule(lr){5-5} \cmidrule(lr){6-8}
Variant & \shortstack{Azimuth\\($^\circ$)} & \shortstack{Elevation\\($^\circ$)} & \shortstack{Distance\\(m)} & \shortstack{AISHELL\\CER (\%)$\downarrow$} & Left/right & \shortstack{Motion\\state} & Trajectory \\
\midrule
Element-wise addition & 13.47 & 7.93 & 1.32 & 5.34 & 91.19 & 82.85 & 64.60 \\
Token interleaving & \textbf{4.79} & \textbf{5.97} & \textbf{1.20} & \textbf{2.81} & \textbf{94.82} & \textbf{87.23} & \textbf{65.49} \\
\bottomrule
\end{tabular*}
\end{minipage}\par\smallskip
\begin{minipage}[t]{0.41\linewidth}
\centering
\setlength{\tabcolsep}{2pt}
\textbf{(b) Temporal correspondence}\par\vspace{3pt}
\begin{tabular*}{\linewidth}[t]{@{\extracolsep{\fill}}lrr@{}}
\toprule
Stream shift & $N$ & Acc. (\%)$\uparrow$ \\
\midrule
None & 16 & 75.00 \\
Spatial only & 16 & 25.00 \\
Both streams & 16 & \textbf{81.25} \\
\bottomrule
\end{tabular*}
\end{minipage}\hfill
\begin{minipage}[t]{0.56\linewidth}
\centering
\setlength{\tabcolsep}{2pt}
\textbf{(c) Temporal coordinates: Acc. (\%)}\par\vspace{3pt}
\begin{tabular*}{\linewidth}[t]{@{\extracolsep{\fill}}lrrr@{}}
\toprule
Task & $N$ & SMRoPE & Sequential \\
\midrule
Time-window direction & 16 & \textbf{81.25} & 12.50 \\
Time-window content & 16 & \textbf{81.25} & 25.00 \\
Untimed direction & 16 & 75.00 & \textbf{81.25} \\
\bottomrule
\end{tabular*}
\end{minipage}
\end{reviewtable}

\textbf{Temporal coordinates.} SMRoPE advances each modality by one position per acoustic time step. Standard sequential positions instead count all interleaved tokens, doubling the within-modality stride and altering positional relations across the entire input, including text. Switching to sequential RoPE at inference degrades time-window direction and content judgments (Table~\ref{tab:design_ablation}c), demonstrating that the learned shared temporal grid facilitates retrieving both sources of evidence at the queried time.

\subsection{Spatial Reasoning and Policy Refinement}
\label{sec:reasoning_refinement}

\textbf{Spatial tasks improved by RL.} Table~\ref{tab:rl_skills} highlights tasks where final-answer rewards enhance policies trained with spatial CoT. Gains in distance estimation and comparison, source relations, and location-conditioned timing link outcome rewards to improved selection and association of spatial observations. Stereo transfer likewise benefits lateral position and motion judgments. Appendix~\ref{app:examples} illustrates corrected decisions under listener rotation and source motion, and Appendix~\ref{app:native_stereo_results} presents supplementary stereo-transfer results.

\begin{reviewtable}
\centering
\small
\caption{Task-level gains after multiple-choice RL. Each supervised policy is compared with its direct RL continuation. Values are TWNM and stereo accuracy or SO-Bench mean per-question score (\%); $\Delta$ is the percentage-point gain before rounding.}
\label{tab:rl_skills}
\setlength{\tabcolsep}{2pt}
\begin{minipage}[t]{0.49\linewidth}
\centering
\begin{tabular*}{\linewidth}[t]{@{\extracolsep{\fill}}lrrrr@{}}
\toprule
Task & $N$ & SFT & RL & $\Delta$ \\
\midrule
\multicolumn{5}{@{}l}{\textit{TWNM / stereo transfer: Acc. (\%)}} \\
Distance estimation & 43 & 41.86 & \textbf{62.79} & +20.93 \\
Distance comparison & 60 & 46.67 & \textbf{51.67} & +5.00 \\
Azimuth direction & 43 & 83.72 & \textbf{86.05} & +2.33 \\
Lateral position (stereo) & 73 & 91.78 & \textbf{93.15} & +1.37 \\
Motion (stereo) & 272 & 37.13 & \textbf{37.87} & +0.74 \\
\bottomrule
\end{tabular*}
\end{minipage}\hfill
\begin{minipage}[t]{0.49\linewidth}
\centering
\begin{tabular*}{\linewidth}[t]{@{\extracolsep{\fill}}lrrrr@{}}
\toprule
Task & $N$ & SFT & RL & $\Delta$ \\
\midrule
\multicolumn{5}{@{}l}{\textit{SO-Bench: Score (\%)}} \\
Distance comparison & 333 & 56.46 & \textbf{57.96} & +1.50 \\
Elevation comparison & 333 & 82.88 & \textbf{84.68} & +1.80 \\
Relative left--right & 333 & 75.98 & \textbf{76.88} & +0.90 \\
Onset from location & 487 & 88.71 & \textbf{89.32} & +0.62 \\
Multi-hop reasoning & 333 & 33.63 & \textbf{33.93} & +0.30 \\
\bottomrule
\end{tabular*}
\end{minipage}
\end{reviewtable}

\section{Conclusion and Future Work}
\label{sec:conclusion}

\ours connects content and geometry through shared input mappings, temporally paired audio and spatial tokens, and SMRoPE. Geometric pretraining, spatial CoT supervision, and outcome rewards support spatial decisions across recording formats. SpheraQA evaluates source attributes, relations, and structured geometry. Results show gains on spatial benchmarks, stereo compatibility, and retention of general-audio capabilities. Future work will strengthen motion and source-relation reasoning across more varied recording conditions. We will release the code, SpheraQA, and evaluation tools.

\label{end:main_text}
\par\pagebreak
\clearpage
\section*{AI Use Statement}
\label{sec:ai_use}

AI tools were used to assist and polish writing, retrieve and discover relevant literature, draft paper sections, and generate synthetic data. The authors take full responsibility for all data, methods, results, conclusions, and the final manuscript.

\section*{Reproducibility Statement}
\label{sec:reproducibility}

The appendix specifies input conventions, model configurations, SMRoPE construction, optimization, and benchmark protocols. The SpheraQA release will include questions, reference annotations, answer-extraction records, and scoring code for reproducing the reported scores.

\bibliographystyle{iclr2027_conference}
{\footnotesize\bibliography{references}}
\clearpage
\appendix
\raggedbottom
\section{Implementation Details}
\label{app:implementation}
\label{app:foa_conventions}

\subsection{Input Features and Sequence Construction}
\label{app:input_features}

FOA represents a sound field with an omnidirectional component $W$ and three directional components in ACN order $[W,Y,Z,X]$. Under SN3D normalization, a plane wave at azimuth $\phi$ and elevation $\theta$ satisfies
\begin{equation}
Y=W\sin\phi\cos\theta,\qquad Z=W\sin\theta,\qquad X=W\cos\phi\cos\theta.
\label{eq:foa}
\end{equation}
For a plane wave $S$ from unit direction $\mathbf u=(u_x,u_y,u_z)$, these relationships yield $[W,Y,Z,X]=S[1,u_y,u_z,u_x]$ and $\operatorname{Re}\{W^*[X,Y,Z]^{\mathsf T}\}=|S|^2\mathbf u$. The cross-spectra between pressure and directional components produce the acoustic pseudo-intensity vector used for localization~\citep{kitic2018tramp}. Relative amplitudes and phases thus provide directional information, while direct-to-reverberant energy provides distance cues.

The input representation uses $16$\,kHz audio in ACN order $[W,Y,Z,X]$ with SN3D normalization. Cartesian axes $X$, $Y$, and $Z$ point forward, left, and upward, respectively. For fully normalized three-dimensional Ambisonics (N3D) recordings, the first-order channels are divided by $\sqrt{3}$. We compute a complex STFT independently for each channel using a $400$-sample Hann window and a $160$-sample hop. Stacking the four real components followed by the four imaginary components yields the $[8,201,T]$ input to Spatial AuT. The audio pathway processes the corresponding omnidirectional signal as a log-mel spectrogram.

The two encoders use the same STFT time grid and corresponding chunk boundaries. Removing the terminal centered-STFT frame in both pathways gives $100$ frames for one second of audio. For a chunk with $T$ frames, three stride-$2$ convolutions produce $K(T)=\lceil T/8\rceil$ feature positions. A complete one-second chunk therefore gives $13$ features per encoder and, after projection, $13$ tokens per pathway. Pairing these output positions before interleaving implements the temporal correspondence in Eq.~\eqref{eq:paired_sequence}.

Table~\ref{tab:input_routing} lists the waveform signals used to construct each encoder's input features. Mono uses the audio pathway alone; stereo uses the mid--side components in Eq.~\eqref{eq:stereo_lift}; and native FOA uses its four components. Higher-order Ambisonics and calibrated microphone arrays enter after external first-order extraction or array-to-Ambisonics encoding~\citep{zotter2019ambisonics,gayer2024ambisonics}. For each complete one-second chunk, projection yields $13$ audio tokens, or $13+13=26$ tokens when the paired spatial pathway is present.

\begin{reviewtable}
\centering
\small
\setlength{\tabcolsep}{3pt}
\renewcommand{\arraystretch}{1.12}
\caption{Input routing and projected token counts. Waveform signals are shown before spectrogram conversion; $a$ denotes the mono waveform. Counts refer to a complete one-second chunk and exclude text and boundary tokens.}
\label{tab:input_routing}
\begin{tabular}{@{}p{0.16\linewidth}p{0.25\linewidth}ccr@{}}
\toprule
Recording & Front-end operation & \shortstack{AuT\\signal} & \shortstack{Spatial AuT\\signals} & Tokens / chunk \\
\midrule
Mono & Audio pathway & $a$ & --- & $13$ \\
Stereo & Mid--side transform & $W_m$ & $[W_m,Y_m,0,0]$ & $26$ \\
FOA & Native first-order components & $W$ & $[W,Y,Z,X]$ & $26$ \\
Higher-order Ambisonics & External first-order extraction & $W$ & $[W,Y,Z,X]$ & $26$ \\
Calibrated array & External array-to-Ambisonics encoding & $W$ & $[W,Y,Z,X]$ & $26$ \\
\bottomrule
\end{tabular}
\end{reviewtable}

\subsection{Encoder and Language Configuration}
\label{app:spatialenc_layers}

The language backbone is initialized from the Qwen3-Omni 4B pretrained LLM. AuT and the audio projector $P_a$ inherit their architectures and pretrained parameters from the same Qwen3-Omni 4B pretrained model, retaining its log-mel front end and temporal downsampling. Spatial AuT is geometrically pretrained, and the spatial projector $P_s$ is learned during subsequent alignment.

Spatial AuT begins with three $3\times3$ convolutional layers, each with stride $2$, padding $1$, and $64$ output channels. The frequency dimension decreases from $201$ to $101$, $51$, and $26$ bins. Flattening the final channel and frequency dimensions produces a $1664$-dimensional vector at each temporal position; a linear layer maps it to the encoder width of $384$. Five temporal attention layers use six heads and a feed-forward width of $1536$. The spatial projector $P_s$ is a two-layer network that uses Gaussian error linear unit (GELU) activations and maps $384\rightarrow384\rightarrow2560$, matching the language embedding width (Table~\ref{tab:architecture_config}).

\begin{reviewtable}
\centering
\small
\caption{Architecture configuration of the spatial encoder and language backbone. FFN denotes the feed-forward network, and KV denotes key/value.}
\label{tab:architecture_config}
\sbox{\reviewtablebox}{%
\begin{tabular}{@{}lrrrrr@{}}
\toprule
Component & Layers & Hidden width & FFN width & Query / KV heads & Head dimension \\
\midrule
Spatial AuT & 5 & 384 & 1536 & 6 / 6 & 64 \\
Language backbone & 36 & 2560 & 9728 & 32 / 8 & 128 \\
\bottomrule
\end{tabular}%
}
\ifdim\wd\reviewtablebox>\linewidth
\resizebox{\linewidth}{!}{\usebox{\reviewtablebox}}
\else\usebox{\reviewtablebox}\fi
\end{reviewtable}

\subsection{Spatial-Encoder Pretraining}
\label{app:spatial_pretraining}

Pretraining attaches six linear prediction heads to the $384$-dimensional Spatial AuT features before the spatial projector. The heads predict azimuth as a sine--cosine pair, elevation, source distance, room floor area, wall absorption, and source activity. Temporal source annotations align these targets with the encoder output positions.

Let $\mathcal V$ contain non-padding positions, $a_i\in\{0,1\}$ denote source activity, and $\mathcal A=\{i\in\mathcal V:a_i=1\}$. For azimuth $\phi_i$ in radians and predicted pair $(\widehat s_i,\widehat c_i)$, the angular loss is
\begin{equation}
\mathcal L_{\mathrm{az}}=\frac{1}{|\mathcal A|}\sum_{i\in\mathcal A}
\left[(\widehat s_i-\sin\phi_i)^2+(\widehat c_i-\cos\phi_i)^2\right].
\label{eq:pretrain_azimuth}
\end{equation}
For continuous attributes $q\in\mathcal Q=\{\mathrm{elevation},\mathrm{distance},\mathrm{area},\mathrm{absorption}\}$, the remaining losses are
\begin{align}
\mathcal L_q&=\frac{1}{|\mathcal A|}\sum_{i\in\mathcal A}(\widehat y_i^q-y_i^q)^2,\label{eq:pretrain_continuous}\\
\mathcal L_{\mathrm{act}}&=-\frac{1}{|\mathcal V|}\sum_{i\in\mathcal V}
\left[a_i\log\sigma(z_i)+(1-a_i)\log(1-\sigma(z_i))\right],\label{eq:pretrain_activity}
\end{align}
where $\widehat y_i^q$ and $y_i^q$ denote the predicted and reference values of attribute $q$, respectively, $z_i$ is the activity logit, and $\sigma$ represents the logistic sigmoid function. Geometry and room losses use active positions; activity supervision includes both sound and silence. Padding is excluded, and a loss with an empty valid set contributes zero. Elevation, distance, and floor area retain units of degrees, meters, and square meters; absorption is dimensionless.

Losses expressed on different numerical scales are balanced using exponential moving averages. For task $k\in\{\mathrm{az},\mathrm{act}\}\cup\mathcal Q$ and optimization step $u$, let $\overline{\mathcal L}_{k,u}$ be its detached loss averaged across data-parallel workers. After initializing $m_{k,u}$ from the first positive observation, we update
\begin{equation}
m_{k,u}=\beta m_{k,u-1}+(1-\beta)\overline{\mathcal L}_{k,u},\qquad
\mathcal L_{\mathrm{pre}}=\sum_k w_k\frac{\mathcal L_{k,u}}{\operatorname{stopgrad}(m_{k,u})}.
\label{eq:pretrain_balancing}
\end{equation}
Here, $\beta=0.99$ denotes the moving-average decay, and $\operatorname{stopgrad}$ treats its argument as a constant during differentiation. The loss weights for azimuth, elevation, distance, floor area, absorption, and activity are set to $(3,3,2,1,1,0.5)$. Moving averages are updated throughout training, omitting tasks with zero loss. This objective trains the encoder to capture temporal geometric features, which the spatial projector subsequently maps into tokens for the language model.

\subsection{SMRoPE: Relative Phase and Causal Pairing}
\label{app:smrope_derivation}

Each $128$-dimensional attention head contains $64$ two-dimensional rotary planes~\citep{roformer}. SMRoPE allocates $24$, $20$, and $20$ such planes to its three position coordinates. Let $\mathcal I_g$ denote the set of plane indices assigned to coordinate $g$, and let $\omega_\ell$ be the rotary frequency of plane $\ell$. For a scalar coordinate $p$, the corresponding block rotation is
\begin{equation}
\mathbf R_g(p)=\operatorname{diag}_{\ell\in\mathcal I_g}
\begin{pmatrix}
\cos(\omega_\ell p)&-\sin(\omega_\ell p)\\
\sin(\omega_\ell p)&\cos(\omega_\ell p)
\end{pmatrix}.
\label{eq:smrope_block_rotation}
\end{equation}
The third coordinate thus acts on $40$ real channels of each head.

The frequency allocation interleaves the three coordinates across the rotary spectrum. With zero-based plane indices, the second and third groups are $\mathcal I_2=\{1+3r:0\leq r<20\}$ and $\mathcal I_3=\{2+3r:0\leq r<20\}$; the first group contains the remaining indices in $\{0,\ldots,63\}$. The rotary base is $10^6$, giving $\omega_\ell=(10^6)^{-\ell/64}$. Temporal and modality-dependent phase shifts are thus applied across multiple frequencies.

\begin{reviewtable}
\centering
\small
\caption{SMRoPE coordinates for the first two spatial--audio token pairs with prefix position $b=0$.}
\label{tab:position_examples}
\sbox{\reviewtablebox}{%
\begin{tabular}{@{}lr@{}}
\toprule
Token & Position coordinates \\
\midrule
$S_1$ & $(1,1,1)$ \\
$A_1$ & $(1,1,2)$ \\
$S_2$ & $(2,2,2)$ \\
$A_2$ & $(2,2,3)$ \\
\bottomrule
\end{tabular}%
}
\ifdim\wd\reviewtablebox>\linewidth
\resizebox{\linewidth}{!}{\usebox{\reviewtablebox}}
\else\usebox{\reviewtablebox}\fi
\end{reviewtable}

For a query at step $i$ with modality $m$ and a key at step $j$ with modality $n$, assigning $0$ to spatial tokens and $1$ to audio tokens gives the key-minus-query coordinates
\begin{equation}
\Delta\mathbf p
=
(j-i,\;j-i,\;j-i+n-m).
\label{eq:relative_phase}
\end{equation}
Let $\mathbf q_i^{(g)}$ and $\mathbf k_j^{(g)}$ denote the query and key components in group $g$. For query modality $m$ and key modality $n$, spatial tokens take $0$ and audio tokens take $1$. Applying $\mathbf R_g(a)^{\mathsf T}\mathbf R_g(b)=\mathbf R_g(b-a)$ to the coordinate assignments in Eq.~\eqref{eq:smrope} yields the rotary attention score
\begin{equation}
\ell_{ij}^{mn}=\frac{1}{\sqrt{128}}\sum_{g=1}^{3}
{\mathbf q_i^{(g)}}^{\mathsf T}
\mathbf R_g\!\left(j-i+\mathbf 1_{g=3}(n-m)\right)
\mathbf k_j^{(g)}.
\label{eq:smrope_attention_kernel}
\end{equation}
The prefix position cancels in the difference. Within each modality, every group follows the temporal displacement $j-i$; across modalities, the third group also incorporates the signed modality displacement $n-m$.

Table~\ref{tab:position_examples} illustrates the coordinates with prefix position $b=0$. In the causal order $S_t,A_t$, an audio query $A_t$ can attend to its paired spatial key $S_t$, yielding relative coordinates $(0,0,-1)$. Its third-group contribution is ${\mathbf q_t^{(3)}}^{\mathsf T}\mathbf R_3(-1)\mathbf k_t^{(3)}/\sqrt{128}$. With the modality coordinates fixed, this bilinear contribution varies with the query and key vectors. The same coordinate construction is retained during prefill and cached decoding.

With prefix position $b$ and $K$ paired time steps as in Eqs.~\eqref{eq:paired_sequence}--\eqref{eq:smrope}, the end-of-audio marker uses $(b+K+1,b+K+1,b+K+1)$. The next text token uses $(b+K+2,b+K+2,b+K+2)$, and subsequent text increments all three axes together. The boundary marker shares its third coordinate with $A_K$, while its first two coordinates advance by one.

\subsection{Optimization}
\label{app:training_details}
\label{app:labels}

Joint SFT combines general-audio instruction instances with spatial question--answer sequences. Spatial targets integrate annotation-derived reasoning traces with final answers, supervising source identities, queried timestamps, spatial observations, and geometric operations. The response loss updates both encoders, both projectors, and the language backbone. Table~\ref{tab:training_config} summarizes the optimization hyperparameters, followed by an illustrative example of the spatial CoT supervision.

\begin{reviewtable}
\centering
\small
\caption{Joint SFT settings.}
\label{tab:training_config}
\sbox{\reviewtablebox}{%
\begin{tabular}{@{}lr@{}}
\toprule
Setting & Value \\
\midrule
Optimizer & AdamW \\
Learning rate & $2\times10^{-6}\rightarrow5\times10^{-7}$ \\
Warmup updates & 100 \\
Global batch size & 256 \\
Training sequence length & 4096 \\
Updates & 1633 \\
Weight decay & 0.1 \\
Adam coefficients & $(0.9,0.95)$ \\
Adam epsilon & $10^{-8}$ \\
\bottomrule
\end{tabular}%
}
\ifdim\wd\reviewtablebox>\linewidth
\resizebox{\linewidth}{!}{\usebox{\reviewtablebox}}
\else\usebox{\reviewtablebox}\fi
\end{reviewtable}

\par\noindent\begin{minipage}{\linewidth}\raggedright
\begin{quote}\small\textbf{Audio-CoT supervision example.} The audio question asks how many sound sources are present. The supervised target is:\par The question asks to count distinct sound sources in this recording.\par Scanning for distinct spatial sources:\par - Source 1: active from 0.00s to 2.50s, to the left (azimuth \textasciitilde{}139°), distance \textasciitilde{}3.5m\par - Source 2: active from 0.00s to 6.75s, to the left (azimuth \textasciitilde{}113°), distance \textasciitilde{}3.0m\par Each source occupies its own (active-window, azimuth, distance) signature. 2 distinct sources identified.\par \par There are 2 sound sources.\end{quote}
\end{minipage}\par\medskip

Reinforcement learning rewards are determined by answer structure and evaluated on the extracted final answer. For numerical answers, the prediction--reference error $e$ is measured as circular distance for azimuth and listener rotation, and as absolute error for elevation and distance. Given the task tolerance $\tau$, the numerical reward is
\begin{equation}
r_{\mathrm{num}}(e;\tau)=
\begin{cases}
1, & e\leq\tau,\\
2-e/\tau, & \tau<e<2\tau,\\
0, & e\geq2\tau.
\end{cases}
\label{eq:rl_numeric_reward}
\end{equation}
For numerical training rewards, the tolerance is $2.5^\circ$ for azimuth, elevation, and listener-rotation answers, and $\max(0.3\,\mathrm{m},0.15d)$ for reference distance $d$. Categorical judgments and multiple-choice answers use $r_{\mathrm{disc}}=\mathbf{1}[\hat c=c^\star]$, where $\hat c$ is the normalized predicted label or option and $c^\star$ is the reference; missing or unparseable predictions receive zero.

Descriptive answers are evaluated against a reference checklist of phrases or numerical entries. Given hit indicators $h_j$ and non-negative weights $w_j$ satisfying $\sum_j w_j>0$, the corresponding reward is
\begin{equation}
r_{\mathrm{list}}=\rho(\hat y)\frac{\sum_j w_jh_j}{\sum_jw_j}.
\label{eq:rl_checklist_reward}
\end{equation}
Here $\hat y$ is the final answer and item weights default to one. Items are matched by normalized phrases, numerical proximity, or word overlap. For a numerical entry $v$, a hit requires an answer value $\hat v$ satisfying $|\hat v-v|<\max(0.5,0.1|v|)$ on the entry's numerical scale. Multiword entries require at least half of their words to appear. When an explicit checklist is absent, entries are drawn from the reference text. The factor $\rho$ is $0.3$ for descriptions shorter than $20$ characters and $1$ otherwise. The checklist measures reference coverage in the final description.

The single RL stage employs three final-answer reward formulations depending on the target format: numerical proximity, categorical or multiple-choice correctness, and hitting-list matching. Optimization updates all modules, with the final supervised policy serving as both the initialization and the fixed KL reference. Table~\ref{tab:rl_config} details the optimization hyperparameters.

\begin{reviewtable}
\centering
\small
\caption{GRPO configuration for the single outcome-based RL stage. The final SFT policy initializes training and remains fixed as the KL reference.}
\label{tab:rl_config}
\begin{tabular}{@{}p{0.62\linewidth}r@{}}
\toprule
Setting & Value \\
\midrule
Updates & $1000$ \\
Questions per update & $32$ \\
Responses per question & $8$ \\
Learning rate (constant) & $5\times10^{-7}$ \\
Adam coefficients & $(0.9,0.999)$ \\
Weight decay & $0.01$ \\
Gradient-norm clip & $1.0$ \\
Policy-ratio clip & $[0.8,1.2]$ \\
Negative-advantage dual clip & $3.0$ \\
KL coefficient & $0.01$ \\
Entropy coefficient & $0.003$ \\
Sampling temperature / top-p & $0.8 / 0.9$ \\
Maximum response length & $256$ tokens \\
\bottomrule
\end{tabular}
\end{reviewtable}

For each group of $G=8$ answers, outcome rewards $r_i$ are centered by their group mean $\bar r$ and scaled by the sample standard deviation $s_r$: $A_i=(r_i-\bar r)/(s_r+10^{-6})$, with $s_r^2=\sum_{i=1}^{G}(r_i-\bar r)^2/(G-1)$. The resulting scalar advantage weights every token in the response. We average token-level losses within each response and then across responses. The policy ratio is clipped to $[0.8,1.2]$, with dual clipping applied at coefficient $3$ for negative advantages. KL regularization employs the low-variance $k_3$ estimator with coefficient $0.01$, and entropy regularization uses coefficient $0.003$.
\section{SpheraQA: Tasks and References}
\label{app:spatialqa}
\label{app:categories}
\label{app:metrics}

SpheraQA contains $20{,}492$ questions across $7{,}059$ recordings, organized into L1 perception, L2 relations, and L3 reasoning. Table~\ref{tab:spatial_main} reports whole-answer accuracy for each question type and the complete benchmark, pooling left--right, front--back, up--down, and near--far as spatial categories and combining the six reasoning question types as Spatial Reasoning. Table~\ref{tab:spatialqa_taxonomy} summarizes the question types, answer formats, and counts throughout the hierarchy.

The recordings total $24.92$ hours. Annotated clip durations have a median of $7.52$ seconds and an interquartile range of $5.21$--$11.44$ seconds. The evaluation covers scalar geometry, source counting, categorical positions, coordinates, motion, trajectories, navigation, source relations, and reasoning that combines spatial observations. Numerical and structured answers complement coarse directional judgments by testing whether the model can express and combine the relevant geometry precisely.

SpheraQA combines simulated static and moving sound scenes, including single and multiple sources, with real recordings from LOCATA~\citep{locata} and STARSS23~\citep{starss23}. Table~\ref{tab:spheraqa_sources} gives the recording and question counts by source. Multichannel evaluation uses the shared four-component spatial representation, while mono-input baselines receive its omnidirectional component.

\begin{reviewtable}
\centering
\small
\caption{Recording sources in SpheraQA. Recordings are counted uniquely within each source, and $N$ denotes the number of questions.}
\label{tab:spheraqa_sources}
\begin{tabular*}{\linewidth}{@{\extracolsep{\fill}}lrr@{}}
\toprule
Source & Recordings & $N$ \\
\midrule
Simulated scenes & 6,828 & 7,047 \\
LOCATA & 63 & 1,140 \\
STARSS23 & 168 & 12,305 \\
\bottomrule
\end{tabular*}
\end{reviewtable}

Questions are constructed from source annotations and structured scene summaries containing source descriptions, activity intervals, positions, trajectories, and room properties. Templates and LLM-assisted generation produce questions and reference explanations at different levels. Each retained question requires an identifiable target and valid references at the queried times. Source evaluation pools and recording identities are preserved; simulated reasoning questions use a held-out scene partition, with question, recording, and scene identities disjoint from the corresponding training and development partitions.

\begin{reviewtable}
\centering
\small
\caption{SpheraQA question types, answer formats, and counts across the L1--L3 hierarchy, corresponding to Table~\ref{tab:spatial_main}.}
\label{tab:spatialqa_taxonomy}
\begin{tabular*}{\linewidth}{@{\extracolsep{\fill}}llr@{}}
\toprule
Question type & Answer format & $N$ \\
\midrule
\multicolumn{3}{@{}l}{\textit{L1 Perception}} \\
Azimuth & Numeric angle & 1,563 \\
Elevation & Numeric angle & 1,556 \\
Distance & Numeric distance & 1,564 \\
Source count & Integer & 1,107 \\
Lateral position & Direction category & 1,315 \\
Motion status & Moving or stationary & 1,604 \\
Motion direction & End--start relation on three axes & 1,314 \\
Spatial categories & Binary spatial attribute & 5,852 \\
Coordinates & Joint coordinate object & 1,451 \\
Trajectory & Time-indexed coordinates & 897 \\
\multicolumn{3}{@{}l}{\textit{L2 Relations}} \\
Distance comparison & Selected source or equal distance & 334 \\
Relative position & Ordered source relation & 229 \\
Navigation & Turn direction & 1,586 \\
\multicolumn{3}{@{}l}{\textit{L3 Reasoning}} \\
Multi-hop reasoning & Source identity and geometric result & 20 \\
Cross-source comparison & Comparison across sources and times & 20 \\
Distance change & Distance-change judgment and reasoning & 20 \\
Acoustic inference & Acoustic property and interpretation & 20 \\
Identity binding & Source identity linked to location & 20 \\
Free-form reasoning & Structured scene interpretation & 20 \\
\midrule
SpheraQA total & & 20,492 \\
\bottomrule
\end{tabular*}
\end{reviewtable}

\begin{reviewtable}
\centering
\small
\caption{SpheraQA evaluation rules. Numerical predictions are scored against reference annotations, and structured outputs must satisfy all required components.}
\label{tab:scoring_rules}
\sbox{\reviewtablebox}{%
\begin{tabular}{@{}p{0.33\linewidth}p{0.58\linewidth}@{}}
\toprule
Criterion & Tolerance or rule \\
\midrule
Azimuth & Circular error $\leq20^\circ$ \\
Elevation & Absolute error $\leq10^\circ$ \\
Distance & Absolute error $\leq1$\,m \\
Coordinates & All requested axes pass \\
Trajectory timestamp matching & $\leq0.051$\,s \\
Trajectory & All required times and coordinates pass \\
Source relations & Match requested attribute and ordered pair \\
\bottomrule
\end{tabular}%
}
\ifdim\wd\reviewtablebox>\linewidth
\resizebox{\linewidth}{!}{\usebox{\reviewtablebox}}
\else\usebox{\reviewtablebox}\fi
\end{reviewtable}

Numerical values and categorical labels are directly parsed into predefined answer fields. For free-form responses, \texttt{gpt-4o-2024-11-20} extracts predicted values and source mentions conditioned solely on the question and model response, without access to ground-truth answers or model identities. Reference-based semantic matching reconciles synonymous source labels and explanatory responses. Deterministic programmatic rules subsequently evaluate geometric tolerances and aggregate all queried components into a whole-answer assessment, applying uniform criteria across all evaluated models. Azimuth uses circular error, elevation uses absolute angular error, and distance uses absolute metric error, with tolerances of $20^\circ$, $10^\circ$, and $1$\,m (Table~\ref{tab:scoring_rules}). Coordinate answers must satisfy all requested axes. Dynamic trajectories must cover all requested times, matched within $0.051$\,s, and satisfy every required coordinate; an explicit static trajectory can specify a single constant position. Categorical answers must match the requested attribute; relational answers must additionally match the ordered source pair.

Each question must have an identifiable source and reference at the required times. Correct answers satisfy all requested fields or axes; missing, unparseable, or incorrect predictions receive zero. Category and overall accuracy use the same fixed question set.

Whole-answer accuracy requires every queried component in a given question to be correct. For reasoning tasks, numerical components adhere to identical angular and distance tolerances as those used in direct localization tasks. Explicitly queried absorption coefficients are evaluated under an absolute tolerance of $0.005$ following fractional conversion from percentages. Any required source identities and explanations must semantically align with the corresponding ground-truth references.

In the single-source fusion study, azimuth MAE is evaluated on $218$ valid predictions per model out of $300$ questions; elevation and distance use
 all $193$ and $253$ questions, respectively. Left--right, motion-state, and trajectory accuracies use all $193$, $274$, and $113$ questions, respectively, treating invalid answers as incorrect. 

\section{Benchmark Protocols}
\label{app:benchmarks}

For SpheraAudio and Spatial-Omni-7B, SpheraQA uses greedy decoding with a $16{,}384$-token limit and shared answer-format prompts. SO-Bench and stereo transfer use $256$ tokens, while stereo-adapted SFT uses $512$. On TWNM, these two models select the option with the highest conditional probability among A--D. SpheraQA reports whole-answer accuracy, SO-Bench averages task-specific per-question scores, and TWNM reports option accuracy. Appendix~\ref{app:prompted_reasoning} provides the standalone explanation-evaluation prompt.

The Gemini baseline uses \texttt{gemini-3.1-pro-preview}. Each request contains a question, its answer-format instruction, and the complete native-audio waveform: FOA $W$ or stereo $(L+R)/2$, at its original sampling rate. Generation uses temperature $0$ and a $16{,}384$-token limit. TWNM scores the generated option letter; SpheraQA follows the extraction and geometric criteria above.

\subsection{Temporal Interventions}
\label{app:temporal_ablations}

Using a fixed SFT checkpoint, we evaluate on 16 confirmation recordings drawn from SO-Bench validation audio in the TAU2019 source family, disjoint at the parent-recording level from a 16-recording development panel. Each full FOA recording provides four-choice questions assessing a named source's nearest horizontal cardinal direction, the sound direction within a specified 0.6-second interval, and the sound category in that interval. Scene annotations identify a unique static target class and an interval containing solely that target. Target directions and answer positions are balanced across the confirmation panel.

We project the audio and spatial tokens once per recording. With $K$ tokens per stream, the stream-shift interventions cyclically shift token vectors by $\lfloor K/2\rfloor$ positions, either within the spatial stream alone or synchronously across both streams to preserve pair associations. Token slots, position IDs, causal masks, boundary tokens, and text remain fixed. For the coordinate intervention, SMRoPE follows Eq.~\eqref{eq:smrope}, whereas sequential RoPE assigns $(i,i,i)$ to physical token index $i$ throughout the complete input, including boundary and text tokens (such that the within-modality stride becomes two instead of one). Token vectors and the causal mask remain unchanged.

All conditions use identical questions and answer keys. Predictions are determined by selecting the option letter among A--D with the highest logit at the first answer position. Each task has $N=16$; paired confidence intervals are computed using 50,000 parent-recording bootstrap resamples stratified by target direction.

For named-source direction, the baseline-minus-spatial-shift and joint-shift-minus-spatial-shift differences are 50.00 [18.75, 81.25] and 56.25 [25.00, 81.25] percentage points, respectively. Under the coordinate intervention, the corresponding SMRoPE-minus-sequential differences are 68.75 [43.75, 93.75] for time-window direction and 56.25 [31.25, 81.25] for content (95\% intervals).

\subsection{SO-Bench and TWNM}
\label{app:spatial_protocol}

SO-Bench combines task-specific scoring rules: localization uses angular or distance tolerances, time-interval queries use interval overlap, and source descriptions and relations use semantic matching. Its aggregate score is the mean of per-question scores computed by task rules, including fractional scores from continuous metrics. For TWNM, we evaluate the final selected option on the fixed question set used in this study, comprising $300$ perception, $300$ relation, and $400$ reasoning questions. The same benchmark-specific rules are applied to all evaluated models.

Spatial prompts specify the queried source, spatial attribute, and answer format. For coordinate queries, SpheraQA requests JavaScript Object Notation (JSON) and states its coordinate convention in the question: the listener is at the origin, azimuth $0^\circ$ points forward, $+90^\circ$ points leftward, and positive elevation points upward. TWNM presents its original options and concludes with \emph{Answer with only one option letter: A, B, C, or D.} SO-Bench retains the task-specific instructions attached to each question. Each evaluation supplies the audio and question in a single user turn.

\par\noindent\begin{minipage}{\linewidth}\raggedright
\begin{quote}\small\textbf{Coordinate prompt.} Provide the spatial coordinates for the sound source in JSON format. Answer in polar coordinates (azimuth in degrees, elevation in degrees, distance in meters). Coordinate system: The listener is at the origin. Azimuth 0$^\circ$ is straight ahead, +90$^\circ$ is to the left, and -90$^\circ$ is to the right; elevation 0$^\circ$ is on the horizontal plane, positive above and negative below; distance is in meters.\par\textbf{Reference answer format.} \texttt{\{"azimuth\_deg": 98.1, "elevation\_deg": 53.2, "distance\_m": 4.44\}}\end{quote}
\end{minipage}\par\medskip

\subsection{General Audio}
\label{app:general_protocol}

The general-audio evaluation suite comprises nine test subsets totaling $69{,}181$ samples. The Base configuration utilizes the Qwen3-Omni 4B pretrained LLM together with its inherited AuT and audio projector. Speech transcription evaluates CER for Mandarin and WER for English under native text normalization. Emotion recognition on IEMOCAP and MELD covers four and seven emotion categories, respectively, while VocalSound assesses six acoustic classes. Classification accuracy is determined by matching canonical labels within normalized outputs, resolving ambiguities in VocalSound by selecting the longest matching class name. Base, Adapted (Mono Data), and Adapted (Mono + Spatial Data) are evaluated on identical audio inputs and ground-truth targets using their respective native prompt templates.

For transcription, the system prompt is \emph{You are a speech recognition model.} English audio uses \emph{Transcribe the English audio into text without any punctuation marks.} The corresponding Chinese instruction asks for plain Chinese text without punctuation. Emotion recognition uses the system prompt \emph{You are a speech emotion recognition model.} followed by \emph{Listen to the audio clip and identify the primary emotion being expressed. Provide keywords.} Vocal-sound classification uses the system prompt \emph{You are a vocal sound classification model.} followed by \emph{Classify the given human vocal sound in English.}

The Adapted (Mono Data) emotion evaluation uses the instruction \emph{Recognize the Meld style emotion with keywords:}. Vocal-sound classification uses the same instruction quoted above.

\subsection{Stereo}
\label{app:stereo_protocol}

BAT estimates source laterality using its direction prompt. For motion questions, BAT compares directional estimates from the two halves of the recording, predicting motion if they differ and falling back to a direct motion query if they coincide. This procedure yields the left--right, lateral-position, and motion predictions reported in Table~\ref{tab:stereo_main}.

\section{Supplementary Spatial Results}
\label{app:extended_results}
\label{app:task_breakdown}

\subsection{SpheraQA}
\label{app:spatialqa_results}

Table~\ref{tab:spatial_main} reports $14$ question groups covering the complete $20{,}492$-question SpheraQA benchmark. The spatial-categories row combines left--right, front--back, up--down, and near--far; Spatial Reasoning pools the six reasoning question types detailed in Table~\ref{tab:spatialqa_taxonomy}. Overall accuracy weights every question equally.

\subsection{SO-Bench}
\label{app:sobench_results}

Table~\ref{tab:sobench_detail} provides supplementary SO-Bench results for source detection, localization, temporal queries, motion, and speech content. These tasks complement the overall benchmark comparison and the policy-refinement analysis presented in the main text.

\begin{reviewtable}
\centering
\small
\caption{Supplementary SO-Bench task scores (\%) for the models in Table~\ref{tab:spatial_main}. Gemini receives the omnidirectional $W$ channel.}
\label{tab:sobench_detail}
\setlength{\tabcolsep}{3pt}
\sbox{\reviewtablebox}{%
\begin{tabular*}{\linewidth}{@{\extracolsep{\fill}}lrrrr@{}}
\toprule
Task & $N$ & \shortstack{Spatial-Omni-7B\\Score (\%)} & \shortstack{Gemini 3.1\\Pro\\Score (\%)} & \shortstack{SpheraAudio\\Score (\%)} \\
\midrule
Source detection & 667 & \textbf{58.77} & 30.88 & 50.07 \\
Time interval & 667 & 83.48 & 48.43 & \textbf{86.96} \\
Azimuth & 1,152 & \textbf{71.44} & 11.37 & 70.05 \\
Elevation & 1,153 & \textbf{76.41} & 32.61 & 70.34 \\
Distance & 486 & \textbf{86.42} & 42.59 & 77.37 \\
Source by direction & 667 & 62.82 & 41.83 & \textbf{66.57} \\
Source by location & 666 & \textbf{59.46} & 32.73 & 58.41 \\
Motion classification & 62 & \textbf{45.16} & 33.87 & 43.55 \\
Source count & 34 & 26.47 & \textbf{44.12} & 38.24 \\
Spatio-temporal query & 333 & 9.88 & 2.10 & \textbf{42.56} \\
Speech content & 171 & 23.39 & 26.32 & \textbf{53.22} \\
\bottomrule
\end{tabular*}%
}
\ifdim\wd\reviewtablebox>\linewidth
\resizebox{\linewidth}{!}{\usebox{\reviewtablebox}}
\else\usebox{\reviewtablebox}\fi
\end{reviewtable}

\subsection{TWNM}
\label{app:twnm_results}

TWNM tests perceptual evidence, source associations, and reasoning about hypothetical scenes. Table~\ref{tab:twnm_levels} reports supplementary task results. Questions involving listener rotation and source removal require maintaining source descriptions while updating the spatial relations under consideration.

\begin{reviewtable}
\centering
\small
\caption{Supplementary TWNM task accuracies (\%). Gemini 3.1 Pro receives only the $W$ channel through its native audio interface. The last column presents the TWNM-SAPO results reported in the original paper~\citep{twnm}. Its evaluation contains $385/279/336$ questions across L1/L2/L3, whereas our evaluation contains $300/300/400$; published values retain their original denominators. The best displayed values in each row are shown in bold.}
\label{tab:twnm_levels}
\setlength{\tabcolsep}{2.5pt}
\sbox{\reviewtablebox}{%
\begin{tabular*}{\linewidth}{@{\extracolsep{\fill}}p{0.29\linewidth}rrrrr@{}}
\toprule
Task category & $N$ & \shortstack{Spatial-\\Omni-7B\\Acc. (\%)} & \shortstack{Gemini\\3.1 Pro\\Acc. (\%)} & \shortstack{SpheraAudio\\Acc. (\%)} & \shortstack{TWNM-SAPO\\(paper)\\Acc. (\%)} \\
\midrule
Source counting & 43 & 55.81 & 62.79 & \textbf{69.77} & 60.00 \\
Environment acoustics & 43 & 48.84 & 46.51 & 62.79 & \textbf{69.09} \\
Event identification & 43 & 65.12 & 74.42 & 69.77 & \textbf{83.64} \\
Speech transcription & 43 & 74.42 & \textbf{97.67} & 81.40 & 60.00 \\
Elevation sign & 42 & 45.24 & 38.10 & \textbf{83.33} & 47.27 \\
\midrule
Identity--location--distance association & 60 & 35.00 & \textbf{78.33} & 66.67 & 71.43 \\
Identity from location & 60 & 48.33 & 55.00 & 65.00 & \textbf{91.07} \\
Location from identity & 60 & 30.00 & 50.00 & \textbf{83.33} & 58.93 \\
Relative direction & 60 & 18.33 & 23.33 & \textbf{63.33} & 45.45 \\
\midrule
Causal intent & 67 & 62.69 & \textbf{92.54} & 88.06 & 82.14 \\
Source-removal counterfactual & 67 & 29.85 & 55.22 & 70.15 & \textbf{78.57} \\
Listener-rotation counterfactual & 67 & 28.36 & 28.36 & \textbf{79.10} & 69.64 \\
Multi-hop query & 67 & 34.33 & 43.28 & 55.22 & \textbf{71.43} \\
Physical consistency & 66 & 80.30 & 53.03 & 84.85 & \textbf{89.29} \\
Scene abduction & 66 & 71.21 & \textbf{92.42} & 89.39 & 87.50 \\
\bottomrule
\end{tabular*}%
}
\ifdim\wd\reviewtablebox>\linewidth
\resizebox{\linewidth}{!}{\usebox{\reviewtablebox}}
\else\usebox{\reviewtablebox}\fi
\end{reviewtable}

Across the three main spatial benchmarks, we estimate uncertainty via a paired bootstrap grouped by source audio. All questions associated with a sampled recording are retained together, and SpheraAudio and Spatial-Omni-7B are evaluated on the identical resampled question set. Table~\ref{tab:paired_uncertainty} presents the $95\%$ bootstrap percentile intervals for question-weighted score differences across $2000$ resamples.

\begin{reviewtable}
\centering
\small
\caption{Source-audio bootstrap $95\%$ confidence intervals for the SpheraAudio minus Spatial-Omni-7B score differences reported in Table~\ref{tab:spatial_main}. Intervals are expressed in percentage points.}
\label{tab:paired_uncertainty}
\sbox{\reviewtablebox}{%
\begin{tabular}{@{}lrr@{}}
\toprule
Benchmark & Audio groups & 95\% interval (pp) \\
\midrule
SpheraQA & 7,059 & [17.17, 20.12] \\
SO-Bench & 7,877 & [1.23, 3.44] \\
TWNM & 961 & [21.82, 29.26] \\
\bottomrule
\end{tabular}%
}
\ifdim\wd\reviewtablebox>\linewidth
\resizebox{\linewidth}{!}{\usebox{\reviewtablebox}}
\else\usebox{\reviewtablebox}\fi
\end{reviewtable}

\section{Supplementary Audio Results}
\label{app:other_audio}

\subsection{Stereo}
\label{app:native_stereo_results}

In stereo transfer, left--right accuracy remains $95.77\%$ across $71$ questions after RL. Table~\ref{tab:rl_skills} reports the lateral-position and motion results for the spatial SFT policy and its RL continuation.

\section{Spatial Question Answering with a Waveform-Analysis Agent}
\label{app:astra_tools}

We evaluate \texttt{gpt-6-astra} with \texttt{xhigh} reasoning effort through Codex on a task-stratified sample of $100$ questions, $25$ from each benchmark. Each question starts in a fresh isolated environment with the public question, anonymous full-duration mono audio, and fixed numerical and plotting tools. The waveform is FOA $W$ or stereo $(L+R)/2$, and the agent analyzes it through waveform and spectrogram tools rather than a native audio encoder.

\begin{reviewtable}
\centering
\small
\caption{Waveform-analysis diagnostic on $25$ selected questions per benchmark.}
\label{tab:astra_tools}
\begin{tabular}{@{}lrlr@{}}
\toprule
Benchmark & $N$ & Metric & Result (\%) \\
\midrule
SpheraQA & 25 & Whole-answer Acc. & 28.00 \\
SO-Bench & 25 & Mean Score & 34.47 \\
TWNM & 25 & Option Acc. & 60.00 \\
Stereo & 25 & Acc. & 0.00 \\
\bottomrule
\end{tabular}
\end{reviewtable}

The cases show useful low-level acoustic analysis but weaker directional and source-time interpretation. In one correctly answered TWNM example, a narrowband peak near $3916$\,Hz and repeated energy pulses identify a beeping sound. All four single-angle questions in the SpheraQA sample instead receive an incorrect $0^\circ$ answer. A temporal query also fails when the observed interval is assigned to the wrong sound category. On stereo questions, $19$ responses abstain and the six submitted choices are incorrect.

\section{Encoder Localization}
\label{app:encoder}
\label{sec:eval_encoder}
\label{app:locata_extended}

The pretrained Spatial AuT localization head evaluates the geometric representation prior to language integration (Table~\ref{tab:encoder_localisation}). It achieves its strongest performance on static speakers and simulated directional estimation. Moving-speaker localization remains more challenging, with steered response power (SRP) performing better on that task and on pooled SELD error. This task profile motivates learning temporally resolved spatial features while retaining direct geometric supervision.

\begin{reviewtable}
\centering
\small
\caption{Localization errors ($^\circ$) of pretrained Spatial AuT and comparison methods. LOCATA evaluates static and moving speakers separately, whereas SELD pools static and moving scenes. The simulated evaluation set contains 223 azimuth and 222 elevation cases. LE denotes localization error; MUSIC is multiple signal classification; PIV is pseudo-intensity vector; DPD is direct-path dominance.}
\label{tab:encoder_localisation}
\sbox{\reviewtablebox}{%
\begin{tabular}{@{}lrrrrr@{}}
\toprule
Model & LOCATA static $\downarrow$ & LOCATA moving $\downarrow$ & SELD LE $\downarrow$ & Sim. azimuth $\downarrow$ & Sim. elevation $\downarrow$ \\
\midrule
Spatial AuT & \textbf{8.23} & 18.25 & 15.78 & \textbf{7.51} & \textbf{3.11} \\
SRP & 9.99 & \textbf{15.24} & \textbf{14.19} & 14.57 & 7.35 \\
MUSIC & 11.12 & 16.64 & 15.49 & 16.12 & 7.46 \\
PIV$+$DPD & 12.54 & 17.95 & 16.62 & 16.96 & 7.90 \\
PSELDNets~\citep{pseldnets} & 18.86 & 46.92 & 40.56 & 16.42 & 8.31 \\
\bottomrule
\end{tabular}%
}
\ifdim\wd\reviewtablebox>\linewidth
\resizebox{\linewidth}{!}{\usebox{\reviewtablebox}}
\else\usebox{\reviewtablebox}\fi
\end{reviewtable}

\section{Spatial Representation Analysis}
\label{app:spatial_analysis}
\label{sec:eval_causal}
\label{app:rotation_full}
\label{app:channel_shuffle_full}

We probe the SpheraAudio supervised policy via controlled transformations of the directional features across $223$ fixed azimuth questions. Yaw rotation acts on the horizontal components as
\begin{equation}
Y'=Y\cos\alpha+X\sin\alpha,\qquad
X'=-Y\sin\alpha+X\cos\alpha,
\label{eq:field_yaw}
\end{equation}
with reference azimuth $\phi'=\operatorname{wrap}(\phi+\alpha)$. Left--right reflection sets $Y$ to $-Y$ and the reference azimuth to $-\phi$. The transforms are applied to the native spatial STFT prior to the encoder. The acoustic features, $W$ component, question tokens, and position indices remain identical across conditions.

\par\medskip\noindent\begin{minipage}{\linewidth}
\expandafter\def\csname @captype\endcsname{figure}
\centering
\includegraphics[width=\textwidth]{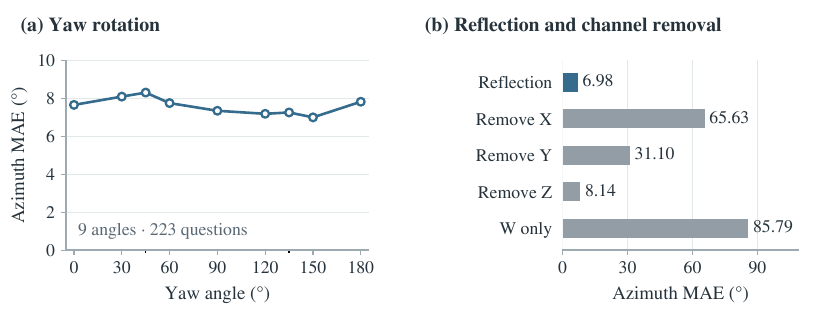}
\caption{Responses to sound-field transformations on 223 fixed azimuth questions. (a) MAE across all nine yaw rotations. (b) Left--right reflection and removal of directional channels. Rotation and reflection use transformed reference directions; channel removal uses original reference directions. The same supervised model is evaluated across all conditions.}
\label{fig:physical_response}
\end{minipage}\par\medskip

On original inputs, the azimuth MAE is $7.66^\circ$. Across the nine yaw angles from $0^\circ$ to $180^\circ$, error against the transformed references remains between $7.00^\circ$ and $8.30^\circ$; left--right reflection yields $6.98^\circ$ (Figure~\ref{fig:physical_response}). The predictions therefore track the prescribed changes in sound-field direction across the tested transformations.

Channel removal identifies the directional features that most strongly influence azimuth estimation. Removing $X$ or $Y$ raises the MAE to $65.63^\circ$ or $31.10^\circ$, respectively, whereas removing $Z$ yields $8.14^\circ$. Retaining only $W$ degrades MAE to $85.79^\circ$. Together with rotation and reflection, these interventions demonstrate that the model has learned to use directional spatial information for localization: its predictions track changes in geometry while the audio-content pathway remains fixed.

\clearpage

\section{Spatial Prompts and Answer Examples}
\label{app:examples}

\subsection{Prompts for Audio-Conditioned Answering}
\label{app:task_prompts}

Each evaluation prompt combines the recording, a task question, and an instruction specifying the answer form. The coordinate convention is given in Appendix~\ref{app:spatial_protocol}. The following boxes pair representative question templates with the corresponding evaluation instructions. Braced fields denote source descriptions, queried times, sampling intervals, or coordinate notes, and are populated during question construction. Numerical questions require numeric values with units, relations identify the queried sources, and temporal questions specify the observations needed over time. For trajectory questions, the answer instruction places the trajectory array inside an object with \texttt{type} and \texttt{trajectory} fields.
\begin{spheracase}[breakable,colframe=teal!55!black,colbacktitle=teal!7!white]{Perception: numerical and categorical answers}
\spherafield{Azimuth question.}{What is the azimuth angle of \{source\_name\}\{time\}?}
\smallskip
\spherafield{Answer instruction.}{Answer with a single signed azimuth value in degrees.}
\smallskip
\spherafield{Coordinate question.}{Provide the spatial coordinates for \{source\_name\}\{time\} in JSON format. \{coord\_note\}}
\smallskip
\spherafield{Answer instruction.}{Return one JSON object with numeric keys azimuth\_deg, elevation\_deg, and distance\_m. Their units are degrees, degrees, and meters.}
\smallskip
\spherafield{Laterality question.}{Is \{source\_name\} on the left side or the right side\{time\}?}
\smallskip
\spherafield{Answer instruction.}{Answer left or right.}
\end{spheracase}
\begin{spheracase}[breakable,colframe=violet!55!black,colbacktitle=violet!7!white]{Source relations and navigation}
\spherafield{Navigation question.}{Which way should I turn to face \{source\_name\}\{time\}?}
\smallskip
\spherafield{Answer instruction.}{Answer with the required turn direction: left, right, turn around, or already facing.}
\smallskip
\spherafield{Distance-comparison question.}{Between \{desc1\} and \{desc2\}, which one is closer to the listener\{time\}?}
\smallskip
\spherafield{Answer instruction.}{Identify which of the two queried sources is closer, or state that they are at the same distance. Use an unambiguous source name or first/second source.}
\smallskip
\spherafield{Relative-position question.}{In terms of left-right positioning, is \{desc1\} to the left or right of \{desc2\}\{time\}?}
\smallskip
\spherafield{Answer instruction.}{Answer the requested spatial relation between the two queried sources. Identify the sources unambiguously using their names or first/second source.}
\end{spheracase}
\begin{spheracase}[breakable,colframe=blue!55!black,colbacktitle=blue!7!white]{Motion and timed trajectories}
\spherafield{Motion question.}{Is \{source\_name\} moving or stationary?}
\smallskip
\spherafield{Answer instruction.}{Answer moving or stationary.}
\smallskip
\spherafield{Trajectory question.}{Describe the spatial trajectory of \{source\_name\} sampled every \{interval\} second(s) as a JSON array. \{coord\_note\}}
\smallskip
\spherafield{Answer instruction.}{Return valid JSON with type (static or dynamic) and trajectory. A source that is stationary over the queried interval needs one position object. For a dynamic trajectory, include every requested timestamp. Each position uses numeric keys azimuth\_deg, elevation\_deg, distance\_m; timed positions also use time\_s. Angles are in degrees, distances in meters, and times in seconds.}
\end{spheracase}
\subsection{LLM-Assisted Training Question Construction}
\label{app:question_generation}

Programmatic CoT construction uses source annotations to instantiate spatial observations and geometric operations, as described in Appendix~\ref{app:training_details}. A complementary LLM-assisted construction path generates varied questions and explanations from a JSON scene summary. This summary contains source descriptions, activity timing, sampled trajectories, and room properties. Each call combines the shared system instruction with the scene summary and one task-specific instruction, and returns separate \texttt{question}, \texttt{cot}, and \texttt{answer} fields. The question is paired with the original recording; the reasoning and final answer form the supervision target. The following excerpts show how the generator is instructed to use scene annotations.
\begin{spheracase}[breakable,colframe=orange!55!black,colbacktitle=orange!7!white]{QA-generation LLM: shared instruction and response schema}
\spherafield{System instruction (excerpt).}{You are an expert spatial audio question-answer generator with deep expertise in 3D sound perception, room acoustics, and auditory scene analysis.\par\smallskip Your task: given a JSON description of a spatial audio scene (with sound source positions, trajectories, room properties), generate ONE high-quality QA pair for training a spatial audio understanding model.}
\smallskip
\spherafield{Response schema (excerpt).}{Return ONLY a valid JSON object with exactly 3 fields:\par - "question": Natural English question. Varied and creative phrasing. If directions/angles are relevant, embed coordinate system definition.\par - "cot": Chain-of-thought reasoning that cites EXACT numerical values from the scene data. Structure: (1) restate what's asked, (2) extract relevant numbers from the provided trajectory data, (3) apply logic/computation step by step, (4) conclude. ONLY reference time points that exist in the trajectory data.\par - "answer": Short final answer (1-2 sentences). Must be verifiable against the metadata.}
\end{spheracase}
\begin{spheracase}[breakable,colframe=orange!55!black,colbacktitle=orange!7!white]{QA-generation LLM: task-specific instruction excerpts}
\spherafield{Numerical direction.}{Generate a question that asks about the **direction** (azimuth) of a specific sound source at a specific time.\par The answerer must read the azimuth value and interpret which side (front/back/left/right) the source is on.\par Embed the coordinate system definition (Azimuth: 0°=ahead, +90°=left, -90°=right, ±180°=behind) in the question.}
\smallskip
\spherafield{Source comparison.}{Generate a question that asks the answerer to **compare two sound sources** — which is closer/further from the listener — at a specific time.\par Embed the coordinate system definition in the question.}
\smallskip
\spherafield{Temporal tracking.}{Generate a **spatial tracking** question about when a dynamic source is in a specific region.\par E.g. "During which time interval is the source saying '...' located to the listener's left?"\par Or: "At what time does the source cross from the right side to the left side of the listener?"}
\smallskip
\spherafield{Listener rotation.}{Generate a **counterfactual** question: "If the listener turns their head X° to the right/left at time T, what would be the new perceived azimuth of source S?"\par The answerer must apply:\par - Turns right: new\_azimuth = old\_azimuth + X\par - Turns left: new\_azimuth = old\_azimuth - X\par Then normalize to [-180, +180]: if > 180, subtract 360; if < -180, add 360.\par Embed the coordinate system definition AND the normalization rule in the question. Use "turns to the right/left" phrasing, NOT "clockwise/counterclockwise".}
\end{spheracase}

\subsection{Model Answer Examples}

The following examples compare numerical localization across models and illustrate subsequent spatial decisions corrected after RL. Each case presents the question, reference, and recorded responses. Quoted model outputs are preserved, with translated source phrases identified where they occur.

A paired response illustrates the temporal decision being refined. When asked whether a designated source crosses from right to left, the supervised policy answers ``No'', whereas its RL continuation correctly identifies the crossing. The reference places the passage behind the listener. Comparing both answers alongside the reference directly reveals this change in spatial judgment.

\begin{spheracase}{A source crossing behind the listener}
\spheraillustratedquery{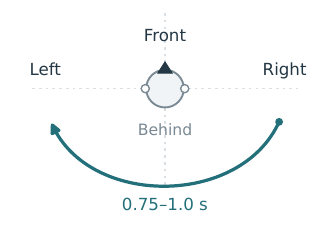}{\spherafield{Reference geometry.}{The source passes from right to left behind the listener between $0.75$ and $1.0$\,s.}\textit{Schematic of the annotated relation.}}
\spheradivider
\spherafield{Question.}{In this scene, does the sound source saying ``Monster Mountain'' ever cross from the listener's right side to their left side during its trajectory? (Azimuth $0^\circ$ is ahead, $+90^\circ$ left, $-90^\circ$ right, and $\pm180^\circ$ behind.)}
\spherarow{SFT.}{No}
\spherarow{RL.}{Yes, during the trajectory, the sound source saying ``Monster Mountain'' crossed from the right side to the left side.}
{\footnotesize\color{SpheraMuted}The source phrase is translated for readability.}
\end{spheracase}

\begin{spheracase}{Numerical localization}
\begin{quote}\small
\textbf{Question (abridged).} Provide the source coordinates in JSON format, using degrees for azimuth and elevation and meters for distance.\par
\textbf{Reference.} \texttt{\{"azimuth\_deg": 98.1, "elevation\_deg": 53.2, "distance\_m": 4.44\}}\par
\textbf{Spatial-Omni-7B.} \texttt{\{"azimuth\_deg": 10.0, "elevation\_deg": 10.0, "distance\_m": 1.8\}}\par
\textbf{SpheraAudio SFT.} \texttt{\{"azimuth\_deg": 96.0, "elevation\_deg": 54.7, "distance\_m": 4.75\}}
\end{quote}
\end{spheracase}

\begin{spheracase}{Source-conditioned localization}
\textbf{Identity followed by localization.} First, identify the adult male speaking rapidly. What is the direction of this voice?\par \textbf{Options.} A) Front-right; B) Front-left; C) Left; D) Back.\par \textbf{Reference.} D.\quad \textbf{SFT.} A.\quad \textbf{RL.} D.
\end{spheracase}

\begin{spheracase}{Listener rotation}
\textbf{Counterfactual listener rotation.} Listen to the laughing and coughing from the crowd. Assuming the coordinate system where azimuth 0 deg = front, 90 deg = left, 180 deg = back, and 270 deg = right, what would be the new azimuth of this sound if the listener turns 180 degrees?\par \textbf{Options.} A) 321.26 degrees (front-right); B) 51.26 degrees (front-left); C) 141.26 degrees (back-left); D) 231.26 degrees (back-right).\par \textbf{Reference.} C.\quad \textbf{SFT.} D.\quad \textbf{RL.} C.
\end{spheracase}

\begin{spheracase}{Following motion over time}
\textbf{Trajectory and crossing time in one real recording.}\par \textbf{Question 1.} How does the source of the blowing sound move throughout the recording?\par \textbf{Options.} A) Laterally from front-right to front-left; B) Straight towards the listener from far away; C) Behind the listener and back to the front; D) Essentially stationary, with changing loudness; E) Laterally from front-left to front-right.\par \textbf{Reference.} E.\quad \textbf{SFT.} A.\quad \textbf{RL.} E.\par \textbf{Question 2.} Approximately when after the start of the recording does the source pass directly in front of the listener (centred left to right)?\par \textbf{Options.} A) 1--2 seconds; B) 3--4 seconds; C) 5--6 seconds; D) 8--9 seconds.\par \textbf{Reference.} C.\quad \textbf{SFT.} B.\quad \textbf{RL.} C.
\end{spheracase}

\subsection{Prompted Spatial Explanations}
\label{app:prompted_reasoning}

The explanation evaluation assesses twelve audio questions, querying the spatial observation, geometric operation, and final option. Greedy decoding generates up to $384$ response tokens. The following examples illustrate a successful listener half-turn alongside a correct option accompanied by inconsistent directional statements.

\begin{spheracase}[breakable,colframe=magenta!45!black,colbacktitle=magenta!5!white]{Explanation-evaluation instruction}
Explain your answer in three short sections:\par 1. Heard spatial evidence: describe the relevant sound source and spatial cues you can hear.\par 2. Geometric steps: connect those observations to the spatial or temporal relation requested in the question.\par 3. Final option: end with a separate line in the form ``Final option: <letter>'', choosing A, B, C, or D.
\end{spheracase}

\begin{spheracase}{Prompted geometric reasoning}
\textbf{A half-turn of the listener.} The harmonica initially lies at $324.35^\circ$ (front-right); after a $180^\circ$ listener turn, the reference is $144.35^\circ$ (back-left), option A.\par
\textbf{SFT.}\par 1. I heard the sound source located at the front-right and below during 0.0s to 5.0s.\par 2. Applying a 180-degree rotation to that position gives back-right and above.\par 3. Final option: D\par
\textbf{RL.}\par 1. I heard the sound source located at the front-right and below.\par 2. Turning the source by 180 degrees would position it at the back-left.\par 3. Final option: A
\end{spheracase}

\begin{spheracase}{Correct option with inconsistent reasoning}
\textbf{A correct option with inconsistent geometry.} The reference changes from $230.11^\circ$ (back-right) to $50.11^\circ$ (front-left), option D, after a $180^\circ$ listener turn. Both policies generate the same response:\par 1. I can hear the sound source located at the back-left position.\par 2. Following the geometric transformation of moving from the back-left to the front-right, the new position is front-left.\par 3. Final option: D
\end{spheracase}

Although the final option is correct, the explanation begins in the wrong quadrant and names two different destination quadrants. This case separates final-answer correctness from the consistency of the generated geometric explanation.

\end{document}